\documentclass[11pt]{article}
\usepackage[a4paper,margin=2.3cm]{geometry}
\usepackage{amsmath,amssymb,bm}
\usepackage{graphicx}
\usepackage{xcolor}
\usepackage[colorlinks=true,linkcolor=blue,citecolor=blue,urlcolor=blue]{hyperref}
\usepackage[font=small,labelfont=bf]{caption}
\usepackage{float}
\usepackage{booktabs}

\newcommand{\dd}{\mathrm{d}}

\title{\bfseries Dephasing-free plasma acceleration \\ with millijoule picosecond lasers}
\author{Alexander Pukhov\\[2pt]
\small Institut f\"ur Theoretische Physik I, Heinrich-Heine-Universit\"at D\"usseldorf, 40225 D\"usseldorf, Germany\\
\small pukhov@tp1.uni-duesseldorf.de}
\date{\today}

\begin{document}
\maketitle

\begin{abstract}
\noindent
Laser-plasma accelerators reach gradients a thousand times above radio-frequency cavities, but need joule-class femtosecond lasers firing about once per second, and electrons outrun the wave. A comb of laser lines spaced by the plasma frequency, each in its own mode of a matched channel, drives a wake moving at the speed of light. In particle-in-cell simulations a 175\,mJ, 0.5\,ps drive holds 21\,GV/m over 10\,mm with a 0.15\,rad phase drift; a test electron gains 208\,MeV, 2.6 times more than in one mode. Because the wave builds up resonantly, the required energy falls steeply with density: a 29\,mJ drive sustains 20\,GV/m, and a 10\,mJ drive, within reach of kilohertz industrial lasers, gives a test electron 170\,MeV over 30\,mm. A simulated stage transfers 5\,\% of the laser energy to the wake; the Manley--Rowe limit is about 18\,\% for five lines.
\end{abstract}

% ------------------------------------------------------------------------------------------------
\section*{Introduction}

Plasma waves driven by intense lasers sustain accelerating fields of 10--100\,GV/m, three orders of magnitude above the breakdown limit of metallic radio-frequency structures~\cite{Tajima1979,Esarey2009}. Laser wakefield accelerators have produced electron beams of several GeV in a few centimetres of plasma~\cite{Gonsalves2019,Picksley2024}. These results rely on femtosecond lasers of petawatt peak power, with tens of joules per pulse and repetition rates of about one shot per second. The proton-driven AWAKE experiment reaches 2\,GeV over 10\,m with a gradient below 1\,GV/m~\cite{AWAKE2018}. Many applications, from compact X-ray and gamma-ray sources to ultrafast electron probes and very-high-energy electron radiotherapy, need the opposite balance: modest energies of 0.1--1\,GeV, but at kilohertz repetition rates and high average power.

The lasers for this exist. Diode-pumped ytterbium thin-disk amplifiers deliver picosecond pulses of 10--200\,mJ at 1--5\,kHz with kilowatt average power and near-perfect beam quality~\cite{Nubbemeyer2017,Dira}. They are, however, the wrong tool for conventional laser wakefield acceleration. A picosecond pulse is about a hundred times longer than the plasma wavelength and cannot excite a wave by itself, and its peak power is far below what is needed for self-guiding. Kilohertz wakefield accelerators driven by millijoule pulses therefore use few-cycle femtosecond lasers in very dense plasma, where the electrons outrun the wave after micrometres to millimetres and gain up to a few tens of MeV~\cite{Faure2019,Salehi2021,Lazzarini2024}.

Two obstacles must be removed to accelerate with a picosecond laser. First, the pulse must be broken up into a train of intensity peaks at the plasma period, so that it excites the plasma wave resonantly, cycle after cycle. This is the idea of the plasma beat-wave accelerator~\cite{Rosenbluth1972,Clayton1993}, of multi-pulse wakefield acceleration~\cite{Hooker2014,Cowley2017} and of the plasma-modulated accelerator~\cite{Jakobsson2021,Ross2023,vandeWetering2023}, in which a picosecond pulse is spectrally modulated into such a train. Second, the wave must not run away from the electrons, or the electrons from the wave. In any guided drive the wave travels at the group velocity of the light, slightly below the speed of light $c$. Relativistic electrons overtake it and are decelerated after the dephasing length. Flying-focus pulses~\cite{Froula2018,Palastro2020,Caizergues2020,Smartsev2019} make the intensity peak travel at $c$ by focusing different parts of the pulse at different places, and traveling-wave schemes~\cite{Debus2019} by overlapping two pulses with tilted fronts at an angle, but they feed the axis from the side and need joule-class lasers, and the fraction of their energy that reaches the wave depends strongly on the pulse duration, the density and the focusing geometry (Supplementary Section~S1). Space-time structured pulses whose intensity peak moves at an arbitrary velocity can be guided in plasma waveguides~\cite{Palastro2025}, and recently Palastro \textit{et al.}~\cite{Palastro2026} showed that a superposition of waveguide modes at frequencies spaced by $\Omega = 2\omega_0(2c/\omega_pw)^2$ forms femtosecond intensity peaks that travel at $c$ and drive a wake without dephasing. In their scheme $\Omega$ is several times smaller than $\omega_p$, each intensity peak drives the wake on its own, and the peak drive amplitudes are $a_0 \approx 0.25$--0.75; in their conclusions they note that recurrences spaced by an integer number of plasma periods would combine resonant wake excitation with dephasing-free acceleration.

Here we combine both requirements in one structure, a \emph{luminal ladder}. We build on our earlier observation that multi-mode beat waves in plasma channels allow the wake velocity to be controlled~\cite{Pukhov2023}. A comb of laser lines is spaced exactly by the plasma frequency $\omega_p$. Line $p$ is sent into the Laguerre--Gaussian mode LG$_p$ of a parabolic plasma channel, and the channel width is chosen such that all lines share one invariant, $\omega - ck_z$. Then every pair of neighbouring lines beats at $\omega_p$ with a phase velocity of exactly $c$: the comb forms a stationary intensity grating in the frame moving with the electrons, and the plasma wave it drives never dephases. The drive is weak ($a_0 \approx 0.02$--0.06 per line; at the beat maxima the on-axis amplitudes add up to $\sum_pa_p \approx 0.10$--0.26 in our simulations), the plasma response linear, and the wave grows resonantly over the whole picosecond pulse. We show with particle-in-cell simulations that a 0.5\,ps, 175\,mJ drive at $3\times10^{18}$\,cm$^{-3}$ holds a gradient of 21\,GV/m over 10\,mm with its phase locked to 0.15\,rad, and that a 29\,mJ drive at twice the density sustains 20\,GV/m. We derive how the required energy scales with density and wavelength, and a simulated 9.8\,mJ drive at $6\times10^{18}$\,cm$^{-3}$ holds 6\,GV/m over 30\,mm; we show where this scaling stops holding, and we give the upper limit of the efficiency. Gradients of this size are today reached with petawatt lasers; here they would come from lasers that run at a kilohertz.

% ------------------------------------------------------------------------------------------------
\section*{Results}

\subsection*{A comb that travels at the speed of light}

\begin{figure}[H]
\centering
\includegraphics[width=\textwidth]{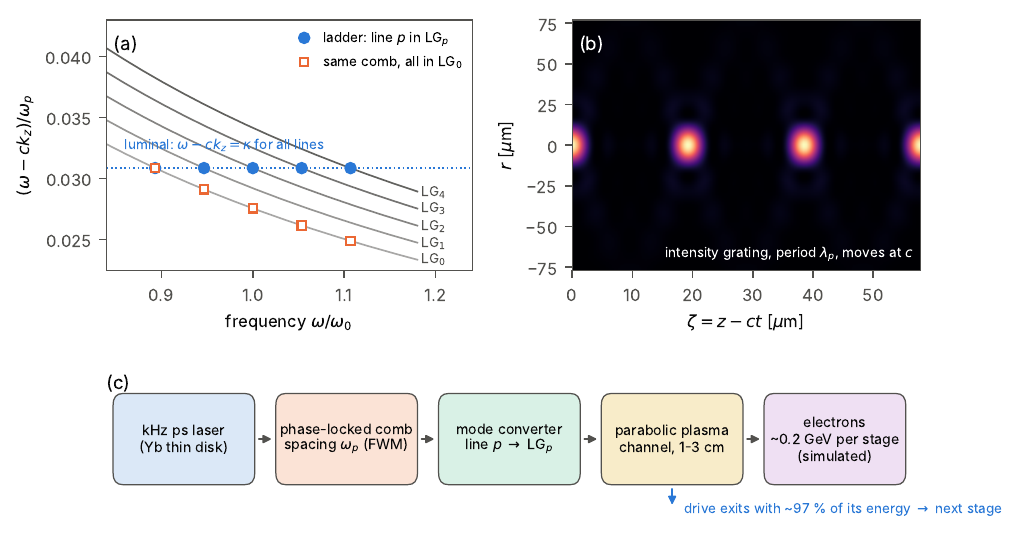}
\caption{\textbf{The luminal ladder.} \textbf{a}, The quantity $\omega - ck_z$ for the guided modes LG$_0$--LG$_4$ of a matched parabolic plasma channel ($n_0 = 3\times10^{18}$\,cm$^{-3}$, $k_pw = 11.4$). Two lines beat at the speed of light if they have the same $\omega - ck_z$. A comb of five lines spaced by the plasma frequency, with line $p$ in mode LG$_p$ (blue dots), lies on one horizontal line: every beat moves at $c$. The same comb with all lines in LG$_0$ (squares) follows the LG$_0$ curve, and its beats are slower than $c$. \textbf{b}, Time-averaged intensity of the five-line ladder in the frame moving at $c$: a grating with the period of the plasma wave that propagates without change. \textbf{c}, Elements of a ladder accelerator.}
\label{fig:concept}
\end{figure}

Light in a plasma channel is guided in discrete transverse modes. In a parabolic channel with density $n(r) = n_0(1 + r^2/R_\mathrm{ch}^2)$ the modes are Laguerre--Gaussian beams LG$_p$ with a matched radius $w$ that does not depend on frequency. Each mode obeys a dispersion relation $c^2k_z^2 = \omega^2 - \omega_p^2 - 4c^2(2p+1)/w^2$, where $k_z$ is the axial wavenumber and $\omega_p$ the plasma frequency on the axis. Two lines beat with the phase velocity $\Delta\omega/\Delta k_z$. This velocity equals $c$ exactly when both lines have the same value of
\begin{equation}
\omega - ck_z = \kappa .
\label{eq:invariant}
\end{equation}
Equation~\eqref{eq:invariant} is the condition for a pattern that is stationary in the co-moving coordinate $\zeta = z - ct$. A flying-focus pulse in free space satisfies it with a continuum of conical waves~\cite{Palastro2020}. In a channel it can be satisfied with a few discrete modes. Figure~\ref{fig:concept}a shows $\omega - ck_z$ for the first five modes. Higher modes have a larger transverse wavenumber, so at a given frequency their $\omega - ck_z$ is larger. A ladder of lines at $\omega_a + p\,\omega_p$, each in LG$_p$, climbs from one mode to the next exactly as fast as $\omega - ck_z$ falls with frequency. This works if
\begin{equation}
\kappa = \frac{4c^2}{w^2\omega_p}
\quad\text{and}\quad
k_p^2w^2 = 8\,\frac{\omega_a}{\omega_p} - 4,
\label{eq:ladder}
\end{equation}
where $k_p = \omega_p/c$ and $\omega_a$ is the lowest line. Equation~\eqref{eq:ladder} holds to first order in $\kappa/\omega$ (Methods). The first condition sets the spacing of the ladder to $\omega_p$; the second fixes the width of the channel. In the simulations the dispersion of every line is computed exactly, and the channel width is fine-tuned so that the residual phase drifts of the neighbour beats, evaluated from the exact dispersion relation, average to zero; the remaining phase motion of the simulated wake is reported below. All lines then lie on one horizontal line in Fig.~\ref{fig:concept}a, and all neighbour beats drive the plasma resonantly and at $c$. In the language of ref.~\cite{Palastro2026}, Eq.~\eqref{eq:ladder} is the special choice $\Omega = \omega_p$ of the luminal mode spacing, the shortest of the resonant recurrence periods proposed in their conclusions. This choice fixes the channel width and turns the superposition into a resonant beat-wave drive; we take it to the limit of a long, weak drive of many plasma periods, which is what allows picosecond pulses of millijoule energy. The time-averaged intensity of the comb is a grating with the period of the plasma wave that does not change as it propagates (Fig.~\ref{fig:concept}b). The same comb with all lines in the fundamental mode LG$_0$, the pulse train of a plasma-modulated or beat-wave accelerator, lies on the LG$_0$ curve instead: its beats are slower than $c$, and its wake slips back relative to the electrons.

The second condition in Eq.~\eqref{eq:ladder} is essential. It ties the channel width to the density and the laser frequency: $k_pw = 11.4$ at $n_0 = 3\times10^{18}$\,cm$^{-3}$ for a ytterbium laser at 1.03\,$\mu$m. A ladder in a channel that misses it drifts in phase and dephases faster than the single-mode comb (Supplementary Section~S5). A second constraint comes from the channel itself: the plasma frequency grows off axis, so the drive stays resonant across the mode only for a limited number of plasma periods $N_\tau = c\tau/\lambda_p$, which requires $k_pw \gtrsim 2\sqrt{N_\tau}$ (Methods).

\subsection*{Dephasing-free acceleration over 10\,mm}

\begin{figure}[H]
\centering
\includegraphics[width=0.92\textwidth]{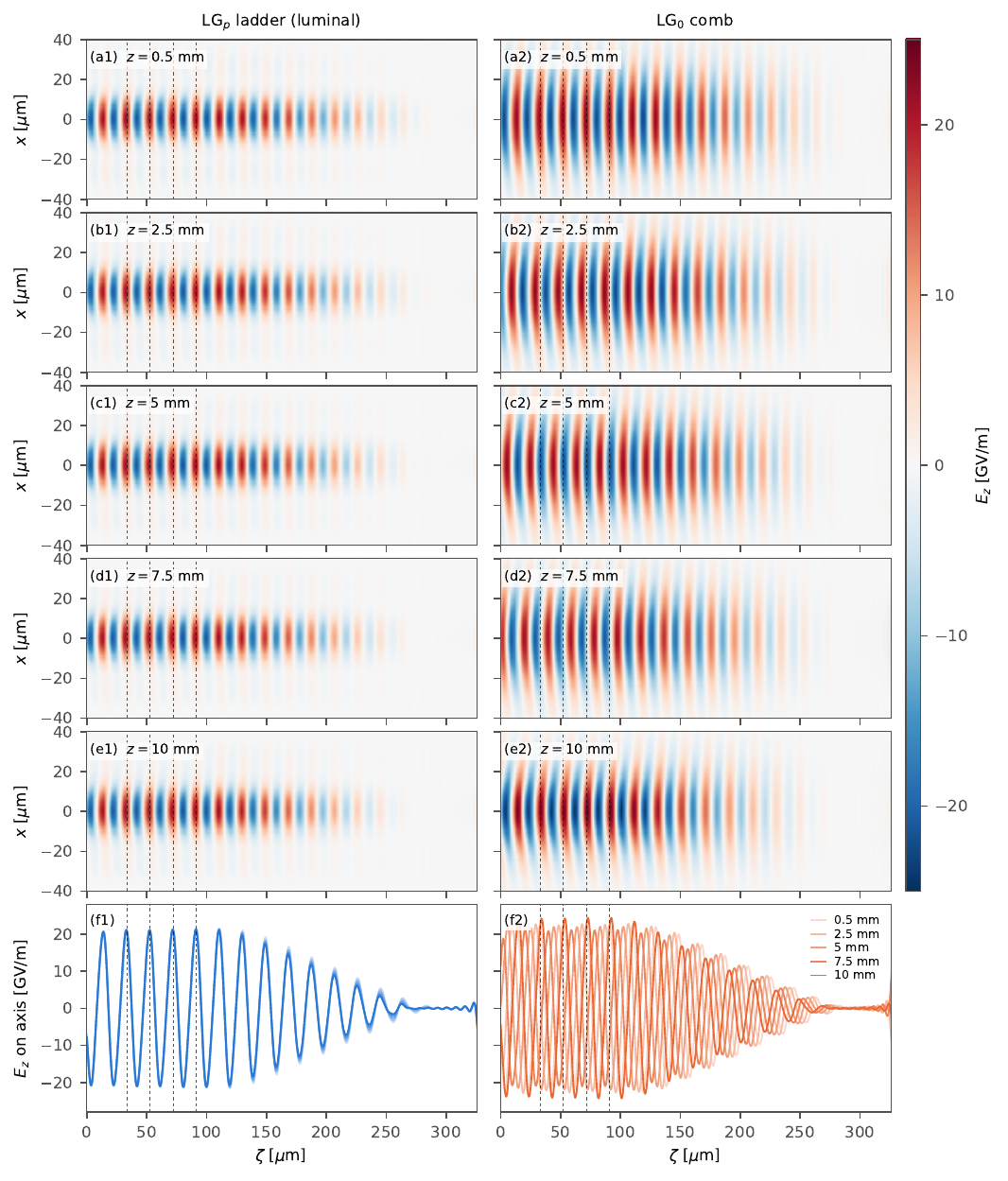}
\caption{\textbf{A plasma wave that does not dephase.} Accelerating field $E_z$ behind the drive in the frame moving at $c$, at five propagation distances from 0.5 to 10\,mm. Left: luminal ladder (five lines, line $p$ in LG$_p$). Right: the same comb with all lines in LG$_0$. Bottom: $E_z$ on the axis at all five distances. The crests of the ladder wave stay in place (dotted lines) and its amplitude stays at 21\,GV/m; those of the single-mode wave drift backwards. 0.5\,ps drive of 175\,mJ, $n_0 = 3\times10^{18}$\,cm$^{-3}$.}
\label{fig:maps}
\end{figure}

\begin{figure}[H]
\centering
\includegraphics[width=\textwidth]{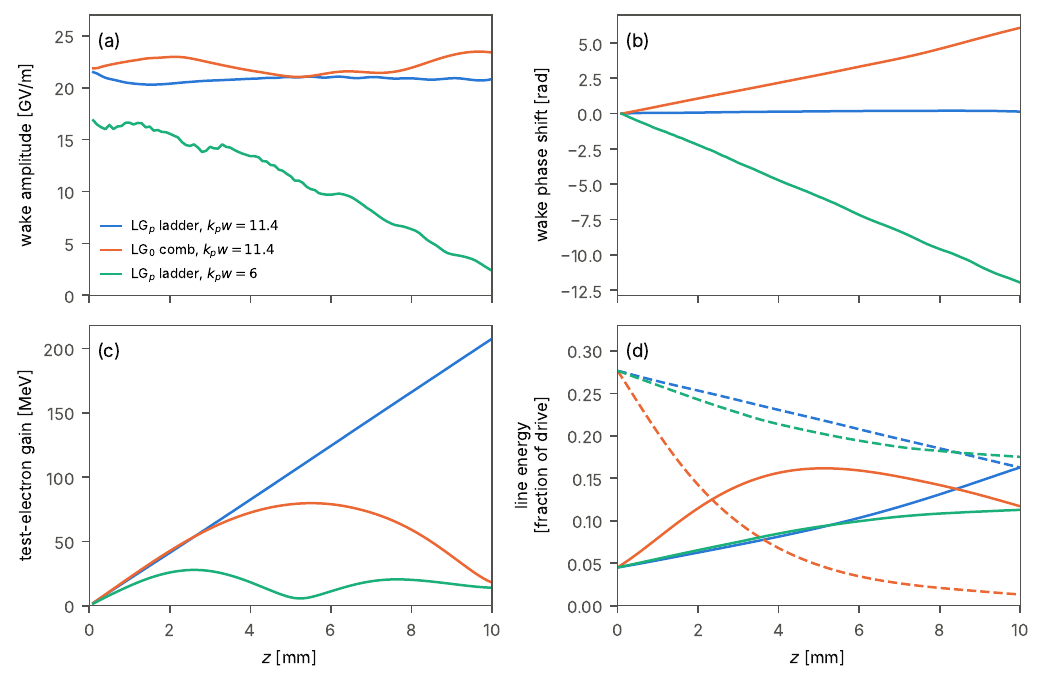}
\caption{\textbf{Locked phase, linear energy gain.} \textbf{a}, Amplitude of the plasma wave on the axis. \textbf{b}, Its phase shift relative to the speed-of-light frame. \textbf{c}, Energy gain of an electron at the best fixed phase. \textbf{d}, Energy of the lowest (solid) and highest (dashed) laser line as a fraction of the drive energy. Blue: luminal ladder; orange: the same comb in LG$_0$; green: a ladder in a channel that misses the width condition, Eq.~\eqref{eq:ladder} ($k_pw = 6$ instead of 11.4).}
\label{fig:ps}
\end{figure}

We simulate the ladder with the particle-in-cell code GEM-PIC-RZ~\cite{GEMPICRZ_paper,GEMPICRZ}, which resolves every optical cycle of every line and computes the propagation of each mode in the plasma, rather than modelling it (Methods). The drive consists of five lines around 1.03\,$\mu$m (1.153--0.931\,$\mu$m), each 0.5\,ps long, with a normalized amplitude $a_0 = 0.057$ (0.028 for the lowest line), in a channel of $n_0 = 3\times10^{18}$\,cm$^{-3}$ that satisfies Eq.~\eqref{eq:ladder}. We compare it with the same comb, at the same energy and in the same channel, but with every line in LG$_0$. We show this low-density case in Figs.~\ref{fig:maps} and \ref{fig:ps} because its wide channel and slow energy flow display the mechanism most clearly; it is not the design point: its backscattering is marginal (below), and the 29\,mJ drive at $6\times10^{18}$\,cm$^{-3}$, with a smaller backscatter estimate, shows the same locked wake against a dephasing single-mode comb (next section and Supplementary Section~S7).

Figure~\ref{fig:maps} shows the plasma wave behind the two drives over 10\,mm. The wave of the ladder does not move: its crests stay at the same position relative to the speed-of-light frame, and its amplitude, 21\,GV/m, is constant (this is the $k_p$ Fourier amplitude of the on-axis field in a window behind the drive; Methods). The wave of the single-mode comb has the same amplitude, but its crests drift steadily backwards. Figure~\ref{fig:ps} quantifies this. The phase of the ladder wave moves by 0.15\,rad over 10\,mm (a linear fit gives 0.015\,rad/mm), that of the single-mode comb by 6\,rad. An electron placed at a fixed phase of the ladder wave gains 208\,MeV over 10\,mm and is still gaining linearly; in the single-mode wake the gain peaks at 80\,MeV after 5.5\,mm and then falls back as the electron enters the decelerating phase. Linear theory predicts 22.5\,GV/m for this drive; the simulated wake is 93\,\% of that value. An amplitude scan at $6\times10^{18}$\,cm$^{-3}$ shows that most of the difference is a weak nonlinearity: the deficit grows with $a_0^2$, as expected from the strong near-axis density modulation of the higher-order mode products, and only about 4\,\% is independent of the amplitude (Supplementary Section~S6).

The drive itself is barely affected inside the channel. The ladder loses 2.6\,\% of its energy over 10\,mm, and the energy of the plasma wave left behind accounts for this loss (Methods). Energy slowly moves down the ladder: each plasma quantum takes a photon from an upper line and returns a photon to the line below (Manley--Rowe), so the top line loses 41\,\% while the bottom line grows, but the sum of the neighbour beats and hence the gradient changes by only 4\,\%. The modes stay pure to better than 98\,\%. The one line that is not on the ladder, the Stokes line below the lowest rung, oscillates with a coherence length of 5\,mm and never takes more than 0.5\,\% of the energy. The single-mode comb, in contrast, cascades down in frequency and loses 12.5\,\% of its energy.

\subsection*{From hundreds to tens of millijoules}

An important question is how much laser energy a given accelerating gradient requires. In the linear regime, the amplitude of the plasma wave grows in proportion to $a_0^2$ and to the pulse length, so a fixed gradient needs a fixed fluence, regardless of whether it is delivered in a shorter, more intense or a longer, weaker pulse. The channel width set by Eq.~\eqref{eq:ladder} then determines the energy. At a fixed gradient $E_z$ it scales as
\begin{equation}
\mathcal{E}_\mathrm{drive} \propto E_z\,\lambda_p^2\,(k_pw)^2\left(\frac{\omega_0}{\omega_p}\right)^2 ,
\label{eq:scaling}
\end{equation}
for a linear plasma response, a matched parabolic channel, and a fixed relative amplitude profile of the lines (Methods). Both $k_pw$ and $\omega_0/\omega_p$ are fixed by Eq.~\eqref{eq:ladder}, $(k_pw)^2 \approx 8\,\omega_0/\omega_p$, so for a given laser $\mathcal{E}_\mathrm{drive} \propto E_z\,n_0^{-5/2}\lambda^{-3}$: doubling the density reduces the energy almost sixfold (Fig.~\ref{fig:scaling}a). The steep fall comes from the channel, which narrows in units of $c/\omega_p$ as the density increases. This is also what limits it: $k_pw$ cannot be made arbitrarily small (below). For a design with the same $k_pw$ the energy falls only as $\lambda_p^2$, and a different laser wavelength merely moves the design to another density, $n_0 \propto \lambda^{-2}$; at the same normalized wake the drive energy then scales as $\lambda$, the gradient as $1/\lambda$ and the stage length as $\lambda$, so a harmonic of the laser saves no energy once the conversion loss is counted. We tested the density scaling at $6\times10^{18}$\,cm$^{-3}$: a 29\,mJ, 0.5\,ps drive, a sixth of the energy of the run above, sustains 16--20.5\,GV/m (0.87--0.90 of linear theory in the first millimetres) over 10\,mm, with a phase slope of 0.047\,rad/mm (0.032\,rad/mm at finer resolution); a test electron at fixed phase gains 181\,MeV and is still gaining at 10\,mm, whereas the same comb in LG$_0$ reaches at most 27\,MeV, after 2\,mm. Energy flows down the ladder faster than at $3\times10^{18}$\,cm$^{-3}$ (the top line drops from 29 to 13\,\% of the drive energy, Fig.~\ref{fig:photons}), and the wake dips to 16\,GV/m at 10\,mm. A drive of a third of this energy, 9.8\,mJ, holds 6.2\,GV/m (0.83--0.87 of linear theory) over 30\,mm with a phase slope of 0.049\,rad/mm, and a test electron gains 169\,MeV. Over 30\,mm this slope adds up to 1.5\,rad, as much as the accelerating-and-focusing phase window, so a real bunch would leave it; because the drift is linear in $z$ and the channel width shifts it by 0.026\,rad/mm per 1\,\% of width, a width offset of 1--2\,\% should null it (projected, not simulated). The two simulations illustrate a general property of the ladder: the wake field and the rate at which energy flows down the ladder both scale as $a^2$, so a weaker drive works over a proportionally longer stage, and the energy gain per stage hardly depends on the drive energy; in both runs the drive gives 5.4\,\% of its energy to the plasma wave before its lines reach the same distribution. The energy is independent of the pulse length, but the resonance is not: the channel constraint $k_pw \gtrsim 2\sqrt{N_\tau}$ limits the duration to $\tau_\mathrm{max} = (k_pw)^2\lambda_p/(4c)$, i.e.\ 2\,ps at $3\times10^{18}$ and 1\,ps at $6\times10^{18}$\,cm$^{-3}$, and longer pulses lose resonance off axis. Raman backscattering, which the simulations do not contain, sets a tighter limit: its gain grows with $E_z\tau$, and our estimate gives a reflectivity of 0.2--2\,\% for the 175\,mJ, 0.5\,ps drive at $3\times10^{18}$\,cm$^{-3}$, $10^{-4}$--$10^{-3}$ for the 29\,mJ drive and below $10^{-5}$ for the 9.8\,mJ drive (Methods). Since energy does not depend on duration, a drive of 22.5\,GV/m is better made shorter than 0.3--0.4\,ps. All our simulations use 0.5\,ps, slightly beyond this estimate. A shorter drive of the same fluence has larger amplitudes, $a \propto \tau^{-1/2}$, and the amplitude scan (Supplementary Section~S6) then gives a larger deficit: for the 29\,mJ drive, 0.85 instead of 0.87 of linear theory at 0.4\,ps and 0.82 at 0.3\,ps (scaled, not simulated). These pulse energies are those of commercial kilohertz thin-disk lasers, which deliver about 0.5--1\,ps; durations of 0.3--0.4\,ps would need some compression.

The scaling does not continue indefinitely. At $10^{19}$\,cm$^{-3}$ it gives 8\,mJ for 22.5\,GV/m, but the simulated 7.8\,mJ, 0.4\,ps drive reaches only 13--15\,GV/m (0.57--0.66 of linear theory), and its phase moves at 0.10\,rad/mm; the result is the same at $\Delta z = 0.5$ and 1\,$\mu$m, so it is not a numerical artefact. The matched channel is narrow at this density ($k_pw = 7.9$, against $k_pw \gtrsim 6.7$ from the detuning criterion for 0.4\,ps), but proximity to the criterion alone does not explain the shortfall: a three-line ladder with a similar margin ($k_pw = 6$ against 5.6) keeps about 0.9 of linear theory (Supplementary Section~S6). The predicted pair drifts are $+0.001$\,rad/mm for all pairs, the near-axis modulation of the mode products is stronger in the narrow channel, and the Stokes line of the lowest rung dephases after only 0.7\,mm; the cause of the shortfall is open. We therefore restrict our projections to $k_pw \gtrsim 9$, i.e.\ $n_0 \le 6\times10^{18}$\,cm$^{-3}$ at 1.03\,$\mu$m, where the simulations follow the scaling.

Higher density is possible here only because there is no dephasing. In a conventional accelerator the dephasing length shrinks steeply with density, and the energy gain with it; this is why multi-GeV stages run at $10^{17}$\,cm$^{-3}$ and need petawatt lasers. The ladder wave moves at $c$ at any density, so the density can be chosen to make the gradient cheap. For comparison, the single-mode comb dephases after 2.3\,mm at $6\times10^{18}$\,cm$^{-3}$ (predicted; simulated: its test-electron gain peaks at 2.0\,mm). The price of high density is precision. At $6\times10^{18}$\,cm$^{-3}$ a density 1\,\% above the design value shifts the wake phase by an additional 0.020\,rad/mm in the simulation (0.022 predicted), and a channel width 1\,\% above it by 0.026\,rad/mm (0.027 predicted); both must therefore be set to about 1\,\% for stages of several millimetres (Methods).

\subsection*{Efficiency}

\begin{figure}[H]
\centering
\includegraphics[width=\textwidth]{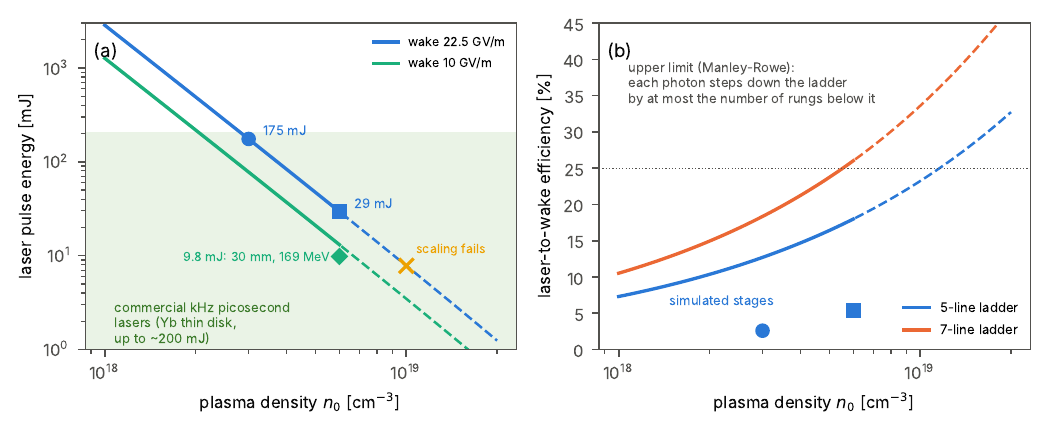}
\caption{\textbf{The laser energy scaling.} \textbf{a}, Laser energy for a linear-theory wake of 22.5 and 10\,GV/m in a five-line ladder at 1.03\,$\mu$m versus plasma density (lines: scaling, Eq.~\eqref{eq:scaling}). The energy does not depend on the pulse length. The duration is limited by the channel-detuning criterion (2.1 and 1.0\,ps at $3\times10^{18}$ and $6\times10^{18}$\,cm$^{-3}$) and, more tightly, by Raman backscattering (estimated: 0.3--0.4\,ps at 22.5\,GV/m; not simulated, Methods). Shaded: pulse energies of commercial kilohertz picosecond lasers. Filled marker: the 10\,mm simulation of Figs.~\ref{fig:maps} and \ref{fig:ps} (simulated wake 21\,GV/m); square: $6\times10^{18}$\,cm$^{-3}$ (simulated over 10\,mm, 16--20.5\,GV/m); diamond: 9.8\,mJ at $6\times10^{18}$\,cm$^{-3}$ (simulated over 30\,mm, 6.2\,GV/m, 169\,MeV); cross: $10^{19}$\,cm$^{-3}$ (simulated, 7.8\,mJ, wake only 13--15\,GV/m: the scaling fails there). Dashed: scaling beyond the simulated range. \textbf{b}, Upper limit (Manley--Rowe) of the fraction of the laser energy that can be transferred to the plasma wave, for five- and seven-line ladders. Markers: the fraction transferred to the wake in simulated stages (circle: 10\,mm at $3\times10^{18}$\,cm$^{-3}$, 175\,mJ; square: $6\times10^{18}$\,cm$^{-3}$, 5.3--5.4\,\% in 10\,mm at 29\,mJ and in 30\,mm at 9.8\,mJ).}
\label{fig:scaling}
\end{figure}

\begin{figure}[H]
\centering
\includegraphics[width=\textwidth]{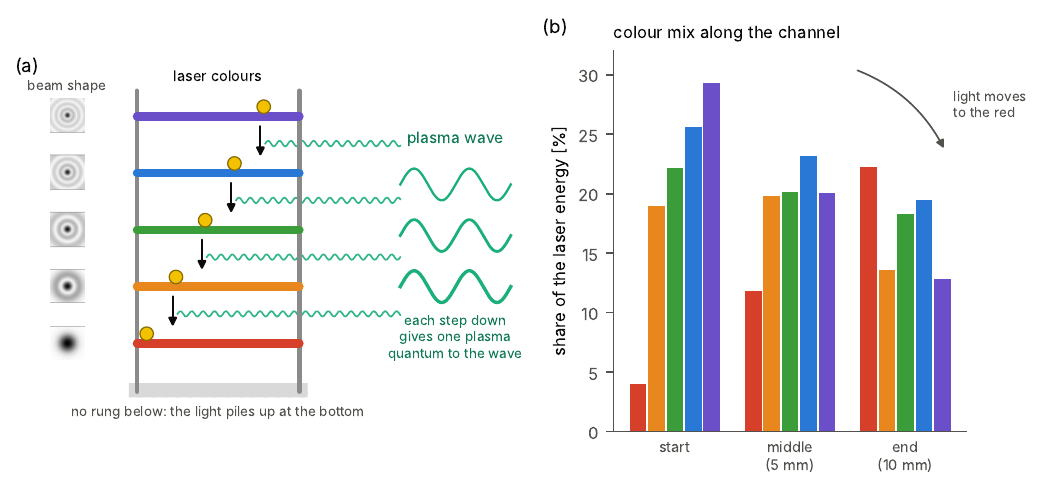}
\caption{\textbf{Upper ceiling for the efficiency (Manley--Rowe).} \textbf{a}, The drive as a ladder: each colour (frequency $\omega_a + q\,\omega_p$) travels in its own beam shape, the guided mode LG$_q$ of the channel. Light gives energy to the plasma wave only by stepping down one rung, one plasma quantum $\hbar\omega_p$ per step; there is no rung below the lowest colour, so the light piles up there. \textbf{b}, Share of the laser energy in each colour at the start, in the middle and at the end of a simulated 10\,mm stage (29\,mJ, $6\times10^{18}$\,cm$^{-3}$); a 9.8\,mJ drive ends in nearly the same distribution after 30\,mm.}
\label{fig:photons}
\end{figure}

The efficiency of a beat-wave drive has a simple upper limit. A photon can transfer energy to the plasma only by stepping down the ladder by one rung, giving up $\hbar\omega_p$, and it cannot step below the lowest rung, whose Stokes line is not guided on the ladder (Fig.~\ref{fig:photons}a). The largest fraction of the drive energy that can go into the plasma is therefore the mean number of rungs below the photons times $\omega_p/\omega$: 12.7\,\% for five lines at $3\times10^{18}$\,cm$^{-3}$, 18\,\% for five lines and 26\,\% for seven lines at $6\times10^{18}$\,cm$^{-3}$, and 23 and 34\,\% at $10^{19}$\,cm$^{-3}$, where, however, the simulated ladder falls short of linear theory (Fig.~\ref{fig:scaling}b). This is an upper bound, not a demonstrated efficiency, and more lines cost some gradient: a simulated seven-line ladder at $6\times10^{18}$\,cm$^{-3}$ (25\,mJ) stays locked (0.029\,rad/mm) but reaches 0.80--0.85 of its linear wake, against 0.87--0.90 for five lines. In the simulations, a 10\,mm stage at $3\times10^{18}$\,cm$^{-3}$ transfers 2.6\,\% of the incident drive energy to the plasma wave; the drive leaves the simulated channel with 97\,\% of its energy, in the same modes and on the same ladder. At $6\times10^{18}$\,cm$^{-3}$ the drive loses 1.4\,\% of its energy in 3\,mm and 5.7\,\% in 10\,mm, of which 94\,\% appears as wake energy (99\,\% at $3\times10^{18}$\,cm$^{-3}$). One stage thus uses about a fifth of the five-line bound at $3\times10^{18}$\,cm$^{-3}$ (2.6 of 12.7\,\%) and about a third at $6\times10^{18}$\,cm$^{-3}$ (5.3--5.4\,\% laser to wake, in 10\,mm at 29\,mJ and 30\,mm at 9.8\,mJ, of 18\,\%; Fig.~\ref{fig:scaling}b). The simulated laser-to-beam efficiencies are far below the bound. At the beam-loading optimum, a 5\,pC witness extracts 20\,\% of the wake energy, i.e.\ 0.11\,\% of the laser energy per millimetre; a 2\,pC witness accelerated over two 10\,mm stages, with the drive of the first stage re-launched by hand into the second, takes 1.1\,\% of the laser energy (Supplementary Section~S8). The limit is the small fraction of its energy, about 0.5\,\% per millimetre, that the linear drive gives to the wave, not the extraction by the beam. Re-using the drive raises the cumulative efficiency, but less than proportionally: the energy that flows down the ladder in the first stage (Fig.~\ref{fig:photons}b) is missing from the upper rungs in the second, which gives two thirds of the first stage's gain; putting more energy into the lowest line of the second stage does not help, because the stronger wake only empties the next line up faster. How much of the drive survives re-timing, mode conversion, transport and coupling in a real staging line is an engineering question that we do not model here.

The ladder also places its energy where it is useful. The single-mode comb deposits 4.7 times more energy per millimetre for the same on-axis field, because its wave is wide; the ladder's wave is concentrated near the axis (Fig.~\ref{fig:maps}), where an electron bunch of micrometre size can extract it.

% ------------------------------------------------------------------------------------------------
\section*{Discussion}

The luminal ladder turns the defining weakness of picosecond lasers, their length, into an advantage. The long pulse builds up the plasma wave resonantly over many periods, the channel modes make every beat travel at the speed of light, and the low intensity keeps the plasma linear and the drive intact. The simulations establish a wake of about 20\,GV/m with a nearly fixed phase from picosecond drives of 175 and 29\,mJ, and a 9.8\,mJ drive holds 6\,GV/m over 30\,mm, with nearly the same energy gain per stage; at $10^{19}$\,cm$^{-3}$ the simulated wake falls well below the scaling, which sets the practical density limit of the present design. Table~\ref{tab:compare} puts this in context, separating simulated from projected values. The closest scheme, the dephasingless accelerator in a plasma waveguide of ref.~\cite{Palastro2026}, works at the same density and wavelength: its ten-mode drive with a peak amplitude $a_0 = 0.5$ gives about 23\,GV/m (their linear estimate), and we estimate from their Eq.~24 that it carries roughly 0.3\,J; its simulated gain per stage, 660\,MeV in 28\,mm, is 3.6 times that of our 10\,mm stage. The ladder sustains 20\,GV/m with 29\,mJ, about ten times less energy, at a peak on-axis amplitude of 0.18: with $\Omega = \omega_p$ every beat drives the plasma resonantly, and for a given laser the energy falls as $n_0^{-5/2}$ (Eq.~\eqref{eq:scaling}).

\begin{table}[H]
\centering
\small\setlength{\tabcolsep}{4pt}
\begin{tabular}{llccc}
\hline
 & gradient & driver & rep. rate & energy gain \\
\hline
proton-driven (AWAKE~\cite{AWAKE2018}) & $<1$\,GV/m & 400\,GeV protons & $\sim$0.03\,Hz & 2\,GeV in 10\,m \\
petawatt LWFA (BELLA~\cite{Gonsalves2019,Picksley2024}) & 30--40\,GV/m & 20--40\,J, 35\,fs & 1\,Hz & 8--10\,GeV \\
kHz LWFA~\cite{Faure2019,Salehi2021,Lazzarini2024} & $\gtrsim$100\,GV/m & 2--26\,mJ, 4--15\,fs & 1\,kHz & up to tens of MeV \\
DLWFA, simulated~\cite{Palastro2026} & $\sim$23\,GV/m$^{\dagger}$ & $\sim$0.3\,J$^{\ddagger}$, 10 modes & -- & 660\,MeV in 28\,mm \\
\hline
ladder, simulated, $3\times10^{18}$\,cm$^{-3}$ & 21\,GV/m & 175\,mJ, 0.5\,ps & -- & 208\,MeV in 10\,mm \\
ladder, simulated, $6\times10^{18}$\,cm$^{-3}$ & 16--20.5\,GV/m & 29\,mJ, 0.5\,ps & -- & 181\,MeV in 10\,mm \\
ladder, simulated, $6\times10^{18}$\,cm$^{-3}$ & 6.2\,GV/m & 9.8\,mJ, 0.5\,ps & (1\,kHz lasers) & 169\,MeV in 30\,mm \\
ladder, simulated, $10^{19}$\,cm$^{-3}$ & 13--15\,GV/m$^{*}$ & 7.8\,mJ, 0.4\,ps & -- & 42\,MeV in 3\,mm \\
\hline
\end{tabular}
\caption{\textbf{Gradient and laser energy.} Upper part: published results (ref.~\cite{Palastro2026}: simulations at $6\times10^{18}$\,cm$^{-3}$ and 1\,$\mu$m, 23\,fs intensity peaks; $^{\dagger}$linear estimate at peak $a_0 = 0.5$, their Eq.~21; $^{\ddagger}$our estimate from their Eq.~24; ref.~\cite{Palastro2026} gives no pulse energy). Lower part: this work; the energy gain is that of a test electron at fixed phase over the simulated length. The ladder drive energies match commercial kilohertz picosecond lasers; the repetition rate is not part of the simulations. $^{*}$0.57--0.66 of linear theory, phase drift 0.10\,rad/mm: the energy scaling does not hold at this density.}
\label{tab:compare}
\end{table}

Three elements are needed to build such an accelerator (Fig.~\ref{fig:concept}c). A candidate route to the comb is cascaded four-wave mixing or molecular modulation in a gas-filled hollow-core fibre or capillary, seeded with two frequencies~\cite{Sokolov2000}. In stimulated rotational Raman scattering in hydrogen the molecule fixes the spacing: the S(1) line, 587\,cm$^{-1}$ (17.6\,THz), equals $\omega_p/2\pi$ at $n_0 = 3.84\times10^{18}$\,cm$^{-3}$, inside the density window of the ladder. The plasma modulator of the plasma-modulated accelerator~\cite{Jakobsson2021,vandeWetering2023} imposes the spacing $\omega_p$ automatically. In such cascades each new line carries the phase $2\varphi_p - \varphi_{p-1}$ of its parents, which is the phase relation the ladder needs; with random line phases the simulated wave falls to 20--80\,\% (Supplementary Section~S4). Whether the required amplitude balance, spacing accuracy, phase stability and conversion efficiency can be reached at tens of millijoules is an open question. The lines, separated by 60--120\,nm, could be sent into their modes with dichroic optics and phase plates, or with a common-path multi-plane light converter; the mode purity required is above about 98\,\%, the value maintained in the simulations. The plasma channel is the most demanding element.

\emph{Practical channels.} The parabolic channel of our simulations is an idealization: an $N$-line ladder needs $N$ guided modes with equally spaced transverse eigenvalues, and a parabola provides this for any number of modes. Real channels are closer to a Gaussian density depression, $n = n_\infty[1 - D\exp(-r^2/s^2)]$, whose depth limits the number of guided modes and whose flattening wings make the mode spacing shrink with mode order. What matters is not the profile itself but the uniformity of the spacing: every line can be placed on its own mode so that each beat moves at $c$, and the beats stay resonant if the spacing is uniform to about $1/(2N_\tau)$, i.e.\ a few per cent for a 0.5\,ps drive. Because the drive energy does not depend on the pulse length, shorter pulses relax this tolerance at no cost in energy. Projecting the ladder onto Gaussian channels (Supplementary Section~S10), a depletion of 20\,\% guides two equally spaced modes, which gives a two-line luminal beat wave~\cite{Pukhov2023} at 1.6 times the energy of the five-line ladder for the same gradient; 30\,\% supports three lines and 40\,\% four lines with spacings uniform to 3--8\,\%. Hydrodynamic optical-field-ionized (HOFI) channels developed in Oxford reach such depths, have been operated at a kilohertz for millions of shots~\cite{Shalloo2018,Alejo2022}, and can be shaped further by conditioning with a second pulse~\cite{Picksley2020}; all-optical channels formed with axiparabolas need only millijoule pulses~\cite{Oubrerie2022}. The ladder requires wider channels ($s \approx 40$--60\,$\mu$m at $6\times10^{18}$\,cm$^{-3}$) than those used so far for guiding femtosecond pulses, and density and width must be set to about 1\,\% (Methods). The number of rungs, and with it the efficiency limit, therefore grows with the depth and quality of the channel: from about 7\,\% for two lines to about 25\,\% for seven lines in a nearly parabolic channel at $6\times10^{18}$\,cm$^{-3}$.

The stage length is not limited by dephasing or by depletion. It is limited by the slow backward slip of the drive envelope relative to the electrons, $c\kappa/\omega$ (3.5\,$\mu$m/mm at $6\times10^{18}$\,cm$^{-3}$), which sets how far behind the drive the electrons must start, and by the slow redistribution of energy along the ladder. The luminal wake ($E_z/E_0 \approx 0.1$) does not trap background electrons, so the ladder is a booster for an external injector, timed to a fraction of $\lambda_p/4$ (3.4\,$\mu$m or 11\,fs at $6\times10^{18}$\,cm$^{-3}$). A few-MeV injector suffices in principle: an electron injected at $\gamma_0$ slips behind $c$ by $1/(2\gamma'\gamma_0)$ with $\gamma' = eE_z/(m_ec^2)$, i.e.\ 1.4\,$\mu$m for $\gamma_0 = 10$ at 18\,GV/m and 2.6\,$\mu$m at 10\,GV/m (projected; our witnesses start at 100\,MeV); synchronizing a separate femtosecond injector with the comb, for example by driving it with a post-compressed part of the same laser, is an open issue. Behind the drive the wake is usable over about 140\,$\mu$m at $6\times10^{18}$\,cm$^{-3}$ (at least 0.76 of the gradient); farther back the off-axis plasma frequency of the channel curves the wave fronts and narrows the accelerating and focusing phase (Supplementary Section~S8). In femtosecond tests of the ladder, electrons and positrons are accelerated alike (Supplementary Section~S3). In the picosecond wake at $6\times10^{18}$\,cm$^{-3}$, a Gaussian witness of 5\,pC is the beam-loading optimum (20\,\% of the wake energy extracted, 1.7\,\% energy spread after 1\,mm); 10\,pC overloads the wake. In a simulated two-stage run with the drive re-used, a 2\,pC witness gains 100\,$\to$\,198\,$\to$\,265\,MeV over 20\,mm and keeps its charge, with an energy spread of 18\,\% and a normalized emittance growing from 0.3 to 0.8\,$\mu$m; neither the bunch length nor the matching were optimized (Supplementary Section~S8). Beams of picocoulomb charge at hundreds of MeV and kilohertz rates would suit compact inverse-Compton sources, electron probes and radiotherapy research.

Several questions remain open: the amplitude-independent part (about 4\,\%) of the difference between the simulated and the linear gradient; the residual phase drift of the wake, which exceeds the predicted spread of the pair drifts and grows with density, only partly as numerical dispersion; the lower ratio to linear theory in the 30\,mm stage (0.83--0.87 against 0.93 from the amplitude scan) and its energy balance, in which only 81\,\% of the drive loss is found as wake energy (Supplementary Section~S7); Raman backscattering, estimated but not simulated (Methods); why the ladder at $10^{19}$\,cm$^{-3}$ falls short of linear theory and whether a different channel or pulse design recovers it; shaped witnesses, matching and energy spread in the picosecond wake; particle-in-cell tests in measured, non-parabolic channel profiles; losses in staging and the rebalancing of the ladder between stages; and the experimental generation of phase-locked, mode-converted combs at the required energies. The luminal ladder offers a route to dephasing-free plasma acceleration at the repetition rates of modern industrial lasers.

% ------------------------------------------------------------------------------------------------
\section*{Methods}

\subsection*{Luminal condition and the width of the channel}
For the matched parabolic channel $n(r) = n_0(1 + r^2/R_\mathrm{ch}^2)$ the guided modes are LG$_p$ with $w^4 = 4R_\mathrm{ch}^2/k_p^2$ and the transverse eigenvalue $k_{\perp,p}^2 = 4(2p+1)/w^2$, independent of frequency. With $k_z = (\omega - \kappa)/c$ the dispersion relation $c^2k_z^2 = \omega^2 - \omega_p^2 - c^2k_{\perp,p}^2$ becomes $2\omega\kappa - \kappa^2 = \omega_p^2 + c^2k_{\perp,p}^2$. To first order in $\kappa/\omega$, the lines $\omega_a + p\,\omega_p$ in LG$_p$ share one $\kappa$ if $(\omega_p^2 + c^2k_{\perp,p}^2)/\omega_p$ grows linearly in $p$ with slope $2\kappa$ and intercept $2\kappa\omega_a/\omega_p$, which gives Eq.~\eqref{eq:ladder}: the slope sets $\kappa = 4c^2/(w^2\omega_p)$, the intercept $k_p^2w^2 + 4 = 8\omega_a/\omega_p$. The residual phase drift of each neighbour beat, $\Delta k_z - k_p$, is computed from the exact dispersion relation; in the simulations $k_pw$ is fine-tuned, by evaluating these drifts for the input parameters before the run, so that their mean vanishes; the residual spread is below 0.002\,rad/mm at $3\times10^{18}$, $6\times10^{18}$ and $10^{19}$\,cm$^{-3}$. The simulated wake phase then moves at 0.015\,rad/mm ($3\times10^{18}$\,cm$^{-3}$) and 0.047\,rad/mm ($6\times10^{18}$\,cm$^{-3}$; fits over 0.3--10\,mm; 0.049\,rad/mm over 30\,mm at 9.8\,mJ); at finer resolution the latter falls to 0.032\,rad/mm, so about a third of it is numerical dispersion of the lines (Supplementary Section~S2). The simulated slopes (0.015, 0.032 at the finer resolution, and 0.10\,rad/mm at $3\times10^{18}$, $6\times10^{18}$ and $10^{19}$\,cm$^{-3}$) exceed the predicted residual spread and grow with density; the part that is neither the predicted spread nor numerical dispersion is not explained. Independent errors: a relative error of 1\,\% in density at fixed spot produces a predicted drift of 0.009, 0.022 and 0.05\,rad/mm at $n_0 = 3\times10^{18}$, $6\times10^{18}$ and $10^{19}$\,cm$^{-3}$ (simulated at $6\times10^{18}$: 0.020\,rad/mm), and 1\,\% in spot size at fixed density 0.010, 0.027 and 0.06\,rad/mm (simulated at $6\times10^{18}$: 0.026\,rad/mm); both errors drift in the same direction and can compensate each other.

\subsection*{Channel-detuning criterion}
Across the mode the local plasma frequency rises by $\Delta\omega_p/\omega_p \approx 2/(k_pw)^2$ at $r = w$. A drive of $N_\tau = c\tau/\lambda_p$ plasma periods excites the plasma resonantly at all radii only if this detuning is smaller than about $1/(2N_\tau)$, i.e.\ $k_pw \gtrsim 2\sqrt{N_\tau}$. A 0.5\,ps drive at $k_pw = 4$ and $3\times10^{18}$\,cm$^{-3}$ ($N_\tau = 7.8$, criterion 5.6) reaches only 54\,\% of the linear wake and falls to 12\,\% within 3\,mm, although its line energies hardly change. Drives that satisfy the criterion follow linear theory (Supplementary Section~S6).

\subsection*{Energy scaling}
For a flat-topped drive of fluence length $L_f$, the linear wake on the axis is $E_z/E_0 = (k_p/4)\sum_p a_pa_{p+1}L_f$ with $E_0 = m_ec\omega_p/e$; the simulations agree with this to 7--13\,\% at $3\times10^{18}$ and $6\times10^{18}$\,cm$^{-3}$ (at $6\times10^{18}$ the deficit is $0.041 + 0.085\,(a/a_\mathrm{ref})^2$ for amplitudes $a$ relative to the 29\,mJ drive, Supplementary Section~S6), but only to 34--43\,\% at $10^{19}$\,cm$^{-3}$. The drive energy is $\sum_p I_p\,\pi w^2L_f/(2c)$ with $I_p \propto a_p^2/\lambda_p^2$, the same for every LG$_p$ of equal on-axis amplitude. Eliminating $a^2L_f$ at fixed $E_z$ gives $\mathcal{E}_\mathrm{drive} \propto (E_z/\omega_p)(c/\omega_p)\,w^2/\lambda^2$, i.e.\ Eq.~\eqref{eq:scaling}; the pulse length cancels. Inserting $w$ from Eq.~\eqref{eq:ladder} gives $\mathcal{E}_\mathrm{drive} \propto E_z\,n_0^{-5/2}\lambda^{-3}$ for a given laser. Assumptions: linear cold-fluid response ($a_0 \ll 1$, density perturbation $\ll 1$), matched parabolic channel satisfying Eq.~\eqref{eq:ladder}, a fixed relative amplitude profile of the lines (lowest line at half amplitude), and a drive longer than the plasma period but within the detuning criterion. The Manley--Rowe limit in Fig.~\ref{fig:scaling}b is $\sum_p\mathcal{E}_p\,p\,\omega_p/\omega_{(p)}\big/\sum_p\mathcal{E}_p$ for the line energies $\mathcal{E}_p$ of a ladder with the lowest line at half amplitude.

\subsection*{Simulations}
GEM-PIC-RZ~\cite{GEMPICRZ_paper,GEMPICRZ,GEMPIC} solves Maxwell's equations in the variables $\zeta = z - ct$ and $s = z$ in quasi-cylindrical geometry (azimuthal modes $m = 0, 1$: the laser field of a linearly polarized drive is $m = 1$, its cycle-averaged intensity has $m = 0$ and $m = 2$ parts; the $m = 2$ part vanishes on the axis; a run with $m = 2$ retained reproduces the wake amplitude, phase slope, drive loss and near-axis density of the $m \le 1$ run to three digits, because the $\cos2\varphi$ part of the cycle-averaged intensity of an $m = 1$ drive enters only through the longitudinal laser field, at order $1/(k_0w)^2 \approx 7\times10^{-5}$). The laser carriers are resolved in $\zeta$ and the slow derivative of the forward field is retained, so that the dispersion of every line in the channel is computed self-consistently. Parameters of the 10\,mm runs: $\Delta\zeta = 0.033\,\mu$m (28 points per wavelength of the highest line), $\Delta r = 0.375\,\mu$m, $\Delta z = 1\,\mu$m; $2\times8$ macroparticles per cell ($r$, $\varphi$) and $8\times32$ inside $r = 4\,\mu$m; mobile protons; open radial boundary with an absorbing layer, the plasma ending inside it. Each line has a super-Gaussian envelope of order 8 in the field and 0.5\,ps intensity FWHM. The flat top is a numerical convenience that shortens the simulation window, not a requirement: for a drive much longer than the plasma period, the linear wake behind the drive depends only on the fluence of the beat source, $\propto\sum_pa_pa_{p+1}\int|f(\zeta)|^2\dd\zeta$, and not on the envelope shape $f$; the femtosecond validation runs used Gaussian envelopes (Supplementary Section~S2). The super-Gaussian envelope (intensity rise time 93\,fs) has a spectral width of 1.6\,THz (FWHM), against 0.85\,THz for a Gaussian of the same duration and 16--28\,THz between neighbouring lines, so the lines are well separated in either case. Convergence was checked over 0.3\,mm in particle number, $\Delta r$ (0.75, 0.5, 0.375\,$\mu$m), $\Delta\zeta$ ($\lambda/28$, $\lambda/36$, $\lambda/40$) and $\Delta z$ (0.5, 1\,$\mu$m); ion species (H, He, immobile) change the wake by less than 1.5\,\% (Supplementary Section~S6). Over the full 10\,mm at $6\times10^{18}$\,cm$^{-3}$, a run with $\Delta r = 0.25\,\mu$m and $\Delta\zeta = \lambda/36$ reproduces the wake amplitude at every $z$, the test-electron gain (181\,MeV in both), the drive loss (5.65 against 5.67\,\%), the line energies and the Stokes energy to within 1\,\%; only the phase slope changes, from 0.047 to 0.032\,rad/mm (Supplementary Section~S8). The first 30--50\,$\mu$m of propagation are a start-up transient of the wake fields and are excluded from all analyses.

\subsection*{Scattering instabilities and collisions}
The simulations do not contain backward-propagating light: the laser carriers are resolved in $\zeta$ and the slow derivative of the forward field is retained, but with $\Delta z = 1\,\mu$m backward waves are not represented, nor is side scattering beyond about $30^\circ$ ($k(1 - \cos\theta)\Delta z > 1$). We therefore estimate these processes. The most important is Raman backscattering of the lines. A backscattered Stokes wave is resonant with one line only, because the neighbouring lines are $\omega_p$ away and the growth rate is small compared with $\omega_p$ ($\gamma/\omega_p = 0.03$--0.12); in the weak-coupling regime $\gamma = (a_p/2)(\omega_p\omega_0)^{1/2}$ for a line of amplitude $a_p$. A plasma element stays under the pump for the duration $\tau$ and the backscattered light crosses it in $\tau/2$, which gives about $N \approx \sqrt2\,\gamma\tau$ amplitude e-foldings, an intensity gain of $e^{2N}$. At a fixed density $a^2\tau \propto E_z$, so $N^2 \propto E_z\tau$. For five lines with the lowest at half amplitude and a flat-top duration of 0.5, 1 and 2\,ps, $N = 8.4$, 12 and 17 at $3\times10^{18}$\,cm$^{-3}$ (175\,mJ), and $N = 7.1$, 10 and 14 (29\,mJ), 5.0, 7 and 10 (14.7\,mJ) and about 4, 6 and 8 (9.8\,mJ) at $6\times10^{18}$\,cm$^{-3}$. Starting from thermal noise ($10^{-10}$--$10^{-9}$ of the pump intensity), the reflectivity is $e^{2N}$ times this, about 1\,\% at $N \approx 8$--9. For the simulated 0.5\,ps drives it is 0.2--2\,\% at 175\,mJ ($N = 8.4$; comparable to the 2.6\,\% transferred to the wake, so this run is marginal), $10^{-4}$--$10^{-3}$ at 29\,mJ ($N = 7.1$), about $10^{-5}$ at 14.7\,mJ and $10^{-6}$ at 9.8\,mJ. A reflectivity below $10^{-4}$ ($N \lesssim 6.5$) at 22.5\,GV/m requires $\tau \lesssim 0.3$\,ps at $3\times10^{18}$ and 0.4\,ps at $6\times10^{18}$\,cm$^{-3}$; since the drive energy does not depend on $\tau$, backscattering favours short drives, and it, rather than channel detuning, sets the longest useful duration. Durations of 1--2\,ps are unsafe at these gradients. These estimates neglect pump depletion and the modulation by the ladder's own wake (below) and are uncertain by the noise level. Light scattered sideways leaves the channel within 50--100\,fs, about one e-folding. Brillouin backscattering, in the strong-coupling regime, grows at $\gamma \approx 2\times10^{12}$\,s$^{-1}$, i.e.\ $N \approx 1$ at 0.5\,ps. Landau damping does not help in a cold channel ($k\lambda_D = 0.12$ at 10\,eV for $k = 2k_0$). The ladder's own wake makes the electrons oscillate at $v/c \approx E_z/E_0 \approx 0.1$, which modulates the phase of the $2k_0$ resonance in the rear half of the drive; whether this suppresses backscattering is unclear, because the sidebands fall on the neighbouring lines. Collisions: at $6\times10^{18}$\,cm$^{-3}$ and 10\,eV the electron--ion collision frequency is $\nu_{ei} \approx 2.6\times10^{12}$\,s$^{-1}$ ($\ln\Lambda \approx 4.7$), and inverse bremsstrahlung absorbs about 0.3\,\% of the drive per centimetre. Collisional damping of a cold wake at the rate $\nu_{ei}/2$ would matter, but the electrons oscillate at $a_pc \approx 0.04c$ in each line and up to $0.1c$ in the wake, far above the thermal velocity $0.004c$, and the collision rate falls as $(v_\mathrm{th}/v)^3 \sim 10^{-3}$: collisions are negligible. A particle-in-cell check of backscattering with a code that resolves backward waves is left for future work.

\subsection*{Diagnostics}
\emph{Wake amplitude and phase:} the complex Fourier component $\hat E_z = (2/L)\int E_z(\zeta)\,e^{-ik_p\zeta}\dd\zeta$ of the on-axis field over the window $\lambda_p < \zeta < \zeta_B - \lambda_p$, where $\zeta_B$ is the rear edge of the drive envelope (field at $10^{-3}$ of its peak) at $z = 0$; for the 10\,mm run at $3\times10^{18}$\,cm$^{-3}$ this is $\zeta = 19$--79\,$\mu$m. At $6\times10^{18}$\,cm$^{-3}$ the drive envelope slips back by $c\kappa/\omega \approx 3.5\,\mu$m/mm (105\,$\mu$m over 30\,mm, against a drive length of 150\,$\mu$m); there the window and the drive-energy cut below follow the centroid of the drive. A window fixed at its $z = 0$ position gives amplitudes within 7\,\% of these, so the decline of the 30\,mm run at its end is not an artefact of the window. The amplitude is $|\hat E_z|$ (not the peak field), the phase $\arg\hat E_z$ relative to its value at the first output; phase slopes are linear fits from $z = 0.3$\,mm. The linear-theory value is $E_0(k_p/4)\sum_pa_pa_{p+1}\int f^2\dd\zeta$ for the input amplitudes. \emph{Test-electron gain:} $|\int\hat E_z(z)\dd z|$, the gain of a test electron at the best fixed phase of the windowed wake. A bunch at fixed $\zeta$ must start far enough behind the drive to stay behind it as the envelope slips (at least 105\,$\mu$m in the 30\,mm run), where the wake is 0.85--0.94 of its value at the drive tail (Supplementary Section~S8); we estimate that this lowers the gain of the 30\,mm run by about 7\,\%. The loaded witnesses of the two-stage run and the beam-loading scan are real bunches at fixed $\zeta$. \emph{Line energies and mode purity:} the field on the plane $\varphi = 0, \pi$ is demodulated at each line frequency and projected onto LG$_0$--LG$_{11}$ of the matched channel ($r < 4w$); the energy of a line is normalized to the total drive energy at $z = 0$, the purity of line $p$ is its LG$_p$ amplitude fraction. \emph{Energy balance:} the drive energy is $\varepsilon_0\int E_x^2\dd V$ ahead of $\zeta_B - 5\,\mu$m. The energy of the plasma wave per unit length is $U_w = (\varepsilon_0/2)\int(|\hat E_z|^2 + |\hat E_r|^2)\,2\pi r\,\dd r$ with $\hat E_r = \partial_r\hat E_z/(ik_p)$; for a linear wave this sum equals the field energy plus the kinetic energy of the oscillating electrons. Over 10\,mm the wake accumulates 0.45\,mJ/mm, 4.53\,mJ in total, against a drive loss of 2.65\,\% (4.6\,mJ).

\subsection*{Use of large language models}
The large language model Claude (Anthropic)~\cite{Claude} was used to set up and run the simulations, reduce the data, make the figures and draft the text under the author's direction; the author checked the derivations and the results and takes full responsibility for the content.

\section*{Data availability}
Input files, reduced data and analysis scripts will be deposited in a public repository (Zenodo) at submission.

\section*{Code availability}
GEM-PIC-RZ is open source~\cite{GEMPICRZ} (release 1.0.0, doi:10.5281/zenodo.23600179).

\section*{Author contributions}
A.P. conceived the scheme, designed and directed the study, and wrote the manuscript.

\section*{Competing interests}
The author declares no competing interests.

% ------------------------------------------------------------------------------------------------
\section*{Acknowledgements}
This work was supported by the Deutsche Forschungsgemeinschaft (DFG, project \mbox{PU~213/6-3}) and the Bundesministerium f\"ur Forschung, Technologie und Raumfahrt (BMFTR, project 05P24PF2). The author gratefully acknowledges the Gauss Centre for Supercomputing e.V. (\url{www.gauss-centre.eu}) for funding this project (lpqed) by providing computing time through the John von Neumann Institute for Computing (NIC) on the GCS Supercomputer JUPITER at J\"ulich Supercomputing Centre (JSC).

% ================================================================================================
% Supplementary Information (from nc_si.tex)
\clearpage
\setcounter{figure}{0}\setcounter{table}{0}\setcounter{equation}{0}\setcounter{section}{0}
\renewcommand{\thefigure}{S\arabic{figure}}\renewcommand{\thetable}{S\arabic{table}}
\renewcommand{\theequation}{S\arabic{equation}}\renewcommand{\thesection}{S\arabic{section}}
\renewcommand{\theHfigure}{S\arabic{figure}}\renewcommand{\theHtable}{S\arabic{table}}
\renewcommand{\theHequation}{S\arabic{equation}}\renewcommand{\theHsection}{S\arabic{section}}
\begin{center}{\Large\bfseries Supplementary Information}\end{center}

% ------------------------------------------------------------------------------------------------
\section{Efficiency of a free-space flying focus}
In the frame $\zeta = z - ct$ an ideal luminal pattern is stationary and every photon conserves $\omega - ck_z = \kappa$. For a luminal Bessel pulse with a fixed cone angle $\theta$ (an axicon-like focus), $\kappa \approx \omega_0\theta^2/2$. Each Bessel ring carries about the energy of the core, and energy at radius $r$ reaches the axis after a distance $r/\theta$. With the core depleting on $L_\mathrm{dep} \sim (\omega_0/\omega_p)^2c\tau$, the fraction of the incident energy that the wake absorbs is of the order of
\begin{equation}
\eta \sim \frac{\lambda}{2\theta^2L_\mathrm{dep}} = \frac{\lambda}{2c\tau}\left(\frac{\omega_p}{\omega_0\theta}\right)^2 .
\end{equation}
It grows for shorter pulses and higher densities: for $\theta = 50$\,mrad, 50\,fs and $n_0 = 10^{18}$\,cm$^{-3}$ at 0.8\,$\mu$m it gives $\eta \approx 0.6\,\%$, whereas for the shorter pulses and higher densities considered for flying-focus accelerators, e.g.\ 20\,fs, $7\times10^{18}$\,cm$^{-3}$, 1\,$\mu$m and 70\,mrad, it gives $\eta \approx 11\,\%$. This is an order-of-magnitude estimate for a fixed cone angle; flying-focus accelerators use axiparabolas, in which the angle of each ray varies with its initial radius, and their efficiency requires a separate calculation. In a channel the same invariant is met by discrete, confined modes; only the slow longitudinal slip $\kappa/\omega$ of the envelope remains.

% ------------------------------------------------------------------------------------------------
\section{Femtosecond validation of the ladder}
\paragraph{Setup.} $\lambda_0 = 0.8\,\mu$m, $n_0 = 1.86\times10^{19}$\,cm$^{-3}$ ($\omega_p/\omega_0 = 0.1033$), $R_\mathrm{ch} = 40.2\,\mu$m, matched $w = 9.95\,\mu$m ($k_pw = 8.07$, which satisfies the width condition for $\omega_a/\omega_p = 8.68$), $\kappa = 6.34\times10^{-3}\omega_0$. Lines LG$_p$ at $\lambda = 0.8922$, 0.8, 0.7251, 0.6630, 0.6107\,$\mu$m, Gaussian envelopes of 20\,$\mu$m ($1/e$), equal on-axis $a_0$ unless stated. $\Delta\zeta = \lambda_0/24$ ($\lambda_0/32$ for the 3\,mm runs), $\Delta r = 0.5\,\mu$m, $\Delta z = 0.5$--1\,$\mu$m.

\begin{figure}[H]
\centering
\includegraphics[width=0.48\textwidth]{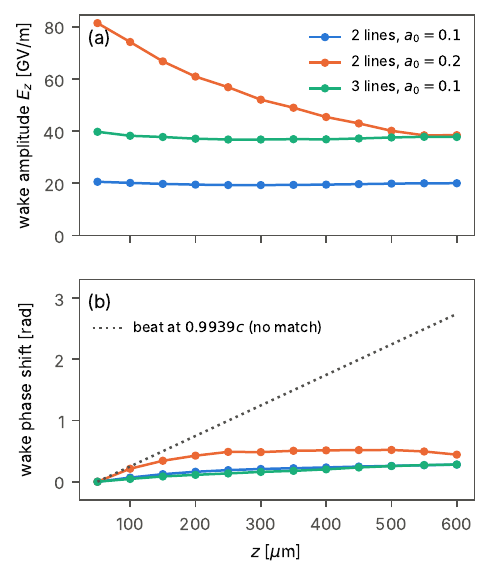}\hfill
\includegraphics[width=0.48\textwidth]{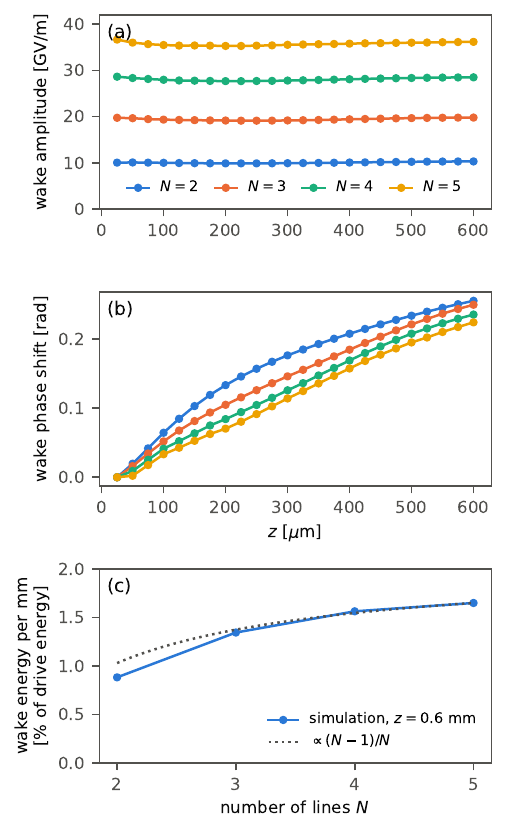}
\caption{\textbf{Femtosecond validation.} Left: wake on the axis behind luminal channel drives, (a) amplitude and (b) phase; dotted: drift of a beat moving at $0.9939c$, the velocity of the same pair without the luminal match. Right: ladders of $N$ lines at $a_0 = 0.07$: (a) wake amplitude, (b) phase, (c) wake energy deposited per mm as a fraction of the drive energy (dotted: $(N-1)/N$).}
\label{fig:wake}\label{fig:ladder}
\end{figure}

\paragraph{Two lines.} Two lines at $a_0 = 0.1$ drive 20\,GV/m, constant to $\pm3\,\%$ over 0.6\,mm. The phase moves by 0.27\,rad, mostly in a start-up of 0.2\,mm, then at $1 - v/c \approx 3\times10^{-4}$, which falls to $2\times10^{-4}$ at $\lambda_0/32$: numerical dispersion of the lines. Without the luminal match the beat would drift by 3\,rad over the same distance (Fig.~\ref{fig:wake}).

\paragraph{Ladders.} Ladders of $N = 2$--5 lines at $a_0 = 0.07$ give 10.3, 19.8, 28.5 and 36.2\,GV/m: the wake grows linearly with the number of beating pairs and stays constant to $\pm2\,\%$; the amplitude per pair falls only from 10.3 to 9.0\,GV/m (Fig.~\ref{fig:ladder}). Every line keeps a mode purity of at least 0.990. The peak on-axis amplitude of the five-line drive is $5a_0 = 0.35$, yet the wake is as clean as that of two lines at $a_0 = 0.1$.

\begin{figure}[H]
\centering
\includegraphics[width=0.48\textwidth]{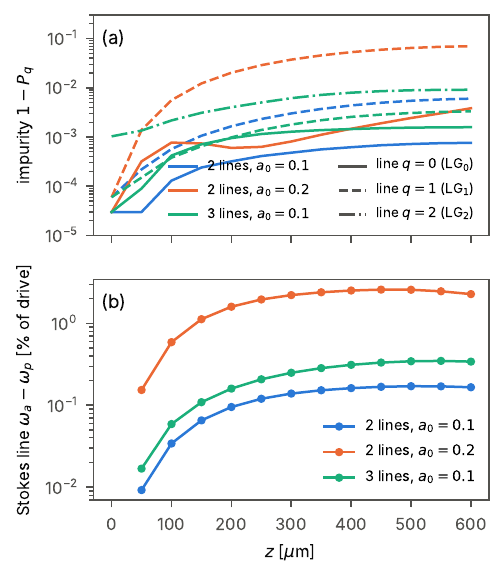}
\caption{(a) Impurity $1 - P_q$ of the drive lines in their modes LG$_q$; (b) energy in the Stokes line $\omega_a - \omega_p$.}
\label{fig:purity}
\end{figure}

\paragraph{The bottom of the ladder.} Forward Raman scattering of the lowest line produces a Stokes line $\omega_a - \omega_p$ that the drive can only put into LG$_0$, off the luminal line. Its beat with $\omega_a$ moves at the uncontrolled velocity and dephases after $\pi/\Delta\kappa$ (0.55\,mm in the fs runs, 5.2\,mm in the ps run of the main text). The anti-Stokes line of the top rung stays on the ladder. At $a_0 = 0.2$ (two lines) the Stokes line takes 2.6\,\% of the drive energy and the wake falls to 0.47 over 0.6\,mm; at $a_0 = 0.1$ it takes 0.17\,\% (Fig.~\ref{fig:purity}). Spreading the drive over more, weaker lines suppresses the leak.

\paragraph{Amplitude profile.} With unequal amplitudes $a_p$ the wake follows $\sum_pa_pa_{p+1}$ to within 3\,\%, and the Stokes energy scales as $a_0^2E_w^2$ with $a_0$ the amplitude of the lowest line (within 20\,\%). The best profile has a weak lowest line and all others at full amplitude: five lines with 0.035 at the bottom and 0.07 above give 32\,GV/m, 13\,\% more than four equal lines, with 3.3 times less Stokes energy.

% ------------------------------------------------------------------------------------------------
\section{Test bunches: electrons and positrons}
\begin{figure}[H]
\centering
\includegraphics[width=0.48\textwidth]{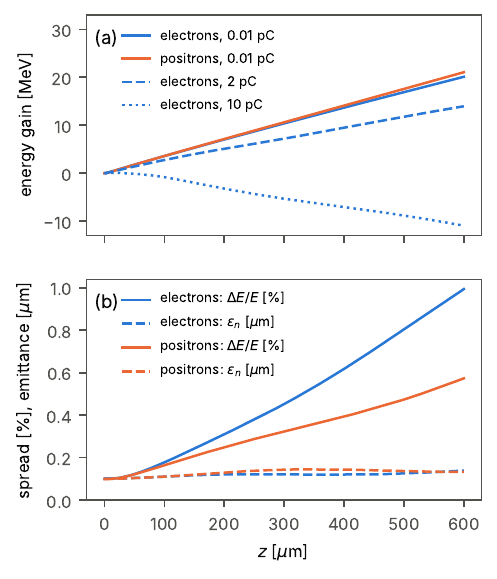}
\caption{Test bunches in the three-line luminal wake ($a_0 = 0.1$): (a) energy gain of passive electron and positron probes (0.01\,pC) and of loaded electron bunches (2 and 10\,pC); (b) energy spread and normalized emittance of the probes.}
\label{fig:witness}
\end{figure}

Electron and positron test bunches (100\,MeV, $\sigma_z = 0.2\,\mu$m, $\sigma_r = 0.5\,\mu$m, $\varepsilon_n = 0.1\,\mu$m) in the three-line fs wake, at the centres of their accelerating-and-focusing quarter periods ($\lambda_p/2$ apart), gain 33.5 and 35.1\,MeV/mm, linearly in $z$ (Fig.~\ref{fig:witness}): there is no dephasing, and electrons and positrons are accelerated alike. A 2\,pC electron bunch (600\,A) loads the wake to 23\,MeV/mm and takes up about 22\,\% of the wake energy; at 10\,pC the bunch overloads the wake. Beam loading of the picosecond wake and a two-stage run: Section~S8.

% ------------------------------------------------------------------------------------------------
\section{Propagation over 3\,mm, Manley--Rowe redistribution and comb phases}
\begin{figure}[H]
\centering
\includegraphics[width=0.48\textwidth]{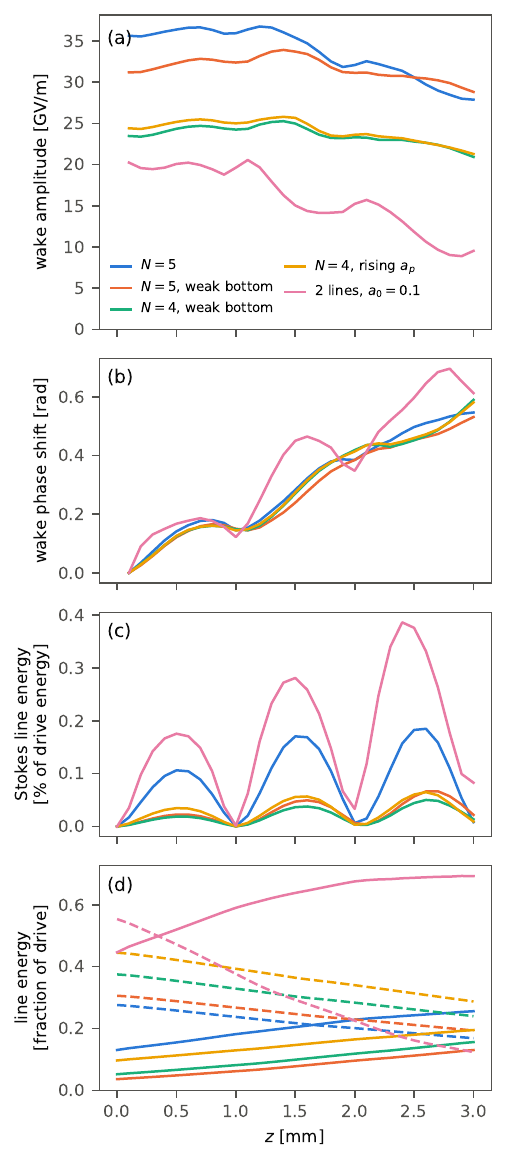}
\caption{Femtosecond drives over 3\,mm: five equal lines ($a_0 = 0.07$), weak-bottom ladders, two lines at $a_0 = 0.1$. (a) Wake amplitude and (b) phase on the axis; (c) energy in the Stokes line; (d) energy fractions of the bottom (solid) and top (dashed) lines.}
\label{fig:long}
\end{figure}

Over 3\,mm (Fig.~\ref{fig:long}) the fs wake is limited not by the Stokes line, which beats with period $2\pi/\Delta\kappa$ and returns to zero, but by Manley--Rowe down-conversion: each plasmon takes a photon from the upper line of a pair and returns one to the lower line, so only $\omega_p/\omega$ of the upper line's loss goes into the wake. The wake predicted from the evolving line energies, $\sum_pa_pa_{p+1}$ with $a_p \propto \mathcal{E}_p^{1/2}/\omega_p$, reproduces the simulated decay to within 3\,\% for the ladders. A weak bottom line is filled by this flow, which is why the tapered ladder keeps its wake longest. In the ps regime the redistribution is much slower (top line $-41\,\%$ over 10\,mm, gradient $-4\,\%$).

\paragraph{Comb phases.} The neighbour beats add in phase only if the line phases satisfy $\varphi_{p+1} - 2\varphi_p + \varphi_{p-1} = 0$. Two ps runs with random line phases give 20 and 79\,\% of the phase-locked wake, with phase shifts of $-2.9$ and $-0.16$\,rad (Table~\ref{tab:controls}). A comb from cascaded four-wave mixing satisfies the condition by construction.

% ------------------------------------------------------------------------------------------------
\section{The width condition of the channel}
A ladder in a channel that misses $k_p^2w^2 = 8\omega_a/\omega_p - 4$ has a different $\kappa$ for every rung. The phase drift of each neighbour beat, $\Delta k_z - k_p$, follows from the exact dispersion relation and agrees with the simulated wake phase in every ps run (Table~\ref{tab:drift}). At $k_pw = 6$ (instead of 11.4) the ladder wake is faster than light by $3.8\times10^{-3}$ and dephases from a witness twice as fast as the single-mode comb; the spread of the pair drifts also dephases the pairs against each other, and the wake falls from 16.8 to 2.4\,GV/m over 10\,mm (Fig.~3 of the main text).

\begin{table}[H]
\centering
\small
\begin{tabular}{lccc}
\toprule
run & $k_pw$ & predicted pair drifts & simulated wake phase slope \\
\midrule
5-line ladder, 48\,mJ & 6 & $-1.43, -1.28, -1.15, -1.04$ & $-1.22$ \\
5-line ladder & 4 & mean $-3.35$ & $-3.25$ \\
3-line ladder & 6 & mean $-1.32$ & $-1.25$ \\
same comb, all LG$_0$ & 6 & $+0.52$ & $+0.62$ \\
5-line ladder, luminal, 175\,mJ & 11.39 & $0.00$ & $+0.018$ \\
same comb, all LG$_0$ & 11.39 & $+0.41$ \ldots $+0.57$ & $+0.60$ \\
\bottomrule
\end{tabular}
\caption{Predicted and simulated wake phase drifts in rad/mm ($n_0 = 3\times10^{18}$\,cm$^{-3}$, 0.5\,ps).}
\label{tab:drift}
\end{table}

% ------------------------------------------------------------------------------------------------
\section{Numerics, convergence and the channel-detuning criterion}
\paragraph{Start-up transient.} In the first 30--50\,$\mu$m of propagation the wake fields jump from zero to their full value; afterwards the simulated wake agrees with linear theory to 1\,\% along the whole ps pulse in a weak two-line test. These outputs are excluded from all analyses.

\paragraph{Near-axis density.} Inside the drive, the electron density near the axis is modulated more strongly than the on-axis $E_z$ suggests. This is linear physics: the transverse Laplacian of the products LG$_p$LG$_{p+1}$ adds 0.4--1.8 times the longitudinal term near the axis, and a local linear cold-fluid model reproduces the simulated modulation at all radii to a few per cent. A deviation behind the drive at coarse radial resolution (a ring at $r \approx 3$--5\,$\mu$m) disappears at $\Delta r = 0.375\,\mu$m, where the density behind the drive also follows the linear model.

\paragraph{Convergence.} Table~\ref{tab:controls} lists the 0.3\,mm controls with the parameters of the main runs. The wake behind the drive changes by less than 2\,\% between $\Delta z = 0.5$ and 1\,$\mu$m, $\lambda/28$ and $\lambda/36$, and for H, He or immobile ions. The ratio to linear theory (0.82 at $k_pw = 6$, 0.90 at $k_pw = 9.3$, 0.93 at $k_pw = 11.4$; 0.90 for three instead of five lines at $k_pw = 6$) does not depend on the numerics. The deficit is largest where the near-axis density modulation is strongest: narrow channels and high-order mode pairs, whose transverse Laplacian adds up to 1.8 times the longitudinal term (see Near-axis density above), so that the near-axis density perturbation reaches about twice the on-axis value of $\delta$. An amplitude scan at $6\times10^{18}$\,cm$^{-3}$ tests this: the five-line ladder of Section~S7 with all amplitudes multiplied by 0.5, 0.71, 1 and 1.41 (7.3--59\,mJ, 1\,mm each) reaches 0.934, 0.918, 0.876 and 0.786 of linear theory (mean over $z = 0.3$--1\,mm; for the 29\,mJ drive the ratio varies between 0.87 and 0.90 over the first 3\,mm, Section~S7). The deficit fits $1 - E_z/E_z^\mathrm{lin} = 0.041 + 0.085\,(a/a_\mathrm{ref})^2$ to within 0.003, with $a_\mathrm{ref}$ the amplitudes of the 29\,mJ drive. About two thirds of the deficit of the 29\,mJ drive therefore scale with $a^2$, as expected from the near-axis nonlinearity; an amplitude-independent part of about 4\,\% remains. Its origin is open; candidates are the difference between the linear model (flat-top fluence, $\sum_pa_pa_{p+1}$) and the actual super-Gaussian drive, and the finite resolution (Section~S8). The near-axis electron density behind the drive grows accordingly, from 0.96--1.02 at the lowest to 0.76--1.36 of the background at the highest amplitude.

\begin{table}[H]
\centering
\begin{tabular}{lcc}
\toprule
run ($k_pw = 6$, 0.3\,mm) & wake / linear theory & wake / reference, phase shift \\
\midrule
reference ($\Delta z = 0.5\,\mu$m, $\lambda/28$, H) & 0.82 & 1, 0 \\
$\Delta z = 1\,\mu$m & 0.82 & 0.998, $+0.001$\,rad \\
$\lambda/36$ & 0.83 & 1.016, $+0.008$\,rad \\
immobile ions & 0.83 & 1.005, $-0.034$\,rad \\
He & 0.83 & 1.014, $-0.005$\,rad \\
same energy in one line & n/a (no beat) & 0.0008, -- \\
random comb phases (1) & 0.17 & 0.20, $-2.94$\,rad \\
random comb phases (2) & 0.64 & 0.79, $-0.16$\,rad \\
\bottomrule
\end{tabular}
\caption{Controls and convergence with the parameters of the main runs ($\Delta r = 0.375\,\mu$m, $2\times8$ macroparticles per cell, $8\times32$ inside 4\,$\mu$m). Wake: $k_p$ Fourier amplitude of the on-axis $E_z$ behind the drive (Methods) at $z = 0.3$\,mm. A single line has no resonant beat, so linear theory predicts no resonant wake; its wake is 0.08\,\% of the reference.}
\label{tab:controls}
\end{table}

\paragraph{Channel-detuning criterion.} The criterion $k_pw \gtrsim 2\sqrt{N_\tau}$ (Methods) was tested deliberately at 0.5\,ps and $3\times10^{18}$\,cm$^{-3}$ ($N_\tau = 7.8$, criterion $k_pw \gtrsim 5.6$): a three-line ladder at $k_pw = 6$, which satisfies it, keeps 0.89--0.92 of the linear wake amplitude over 3\,mm, whereas a five-line ladder at $k_pw = 4$, which violates it, reaches only 0.54 and falls to 0.12 within 3\,mm, although its line energies hardly change. The loss is the off-axis plasma falling out of resonance, not depletion. (Both runs also miss the width condition and drift in phase, Table~\ref{tab:drift}; the criterion concerns the amplitude.) The main runs satisfy the criterion with margin ($k_pw = 11.4$ and 9.3 against 5.6 and 6.6). A seven-line ladder at $k_pw = 6$ decays to 0.53 with near-axis spikes; it misses the width condition. Its luminal version at $6\times10^{18}$\,cm$^{-3}$ ($k_pw = 8.8$, 25\,mJ, 3\,mm) stays locked (phase slope 0.029\,rad/mm) but reaches only 0.80--0.85 of linear theory, against 0.87--0.90 for five lines: more lines raise the Manley--Rowe limit but cost some gradient.

\paragraph{$10^{19}$\,cm$^{-3}$.} A five-line ladder at $10^{19}$\,cm$^{-3}$ ($k_pw = 7.86$, $w = 13\,\mu$m, 0.4\,ps, 7.8\,mJ, linear-theory wake 22.5\,GV/m) reaches 13--15\,GV/m (0.57--0.66 of linear theory) over 3\,mm, with a phase slope of 0.10\,rad/mm, a test-electron gain of 42\,MeV and a drive loss of 4.7\,\%; the near-axis density stays within 0.90--1.06. Runs at $\Delta z = 0.5$ and 1\,$\mu$m give the same wake and slope to three digits, so the shortfall is physical. The matched channel is narrow ($k_pw = 7.9$ against 6.7 from the detuning criterion for $N_\tau = 11.4$), but proximity to the criterion alone does not explain the shortfall: the three-line ladder above, with a similar margin ($k_pw = 6$ against 5.6), keeps 0.89--0.92 of linear theory. The predicted pair drifts, from the exact dispersion for the input parameters, are $+0.001$\,rad/mm for all four pairs, a hundred times below the simulated phase slope. The cause is open; the energy scaling of the main text is therefore restricted to $n_0 \le 6\times10^{18}$\,cm$^{-3}$.

% ------------------------------------------------------------------------------------------------
\section{The ladder at $6\times10^{18}$\,cm$^{-3}$ and the tolerance to density errors}
A five-line ladder at $6\times10^{18}$\,cm$^{-3}$ in the luminal channel ($k_pw = 9.257$, $w = 20\,\mu$m), 0.5\,ps, $a_0 = 0.020$ (lowest line) and 0.040, carries 29\,mJ (linear-theory wake 22.5\,GV/m). Over the simulated 3\,mm the wake stays at 19.5--20.3\,GV/m (0.87--0.90 of linear theory), the phase moves at 0.038\,rad/mm, a test electron at fixed phase gains 60\,MeV, the drive loses 1.4\,\% of its field energy, and the Stokes line of the lowest rung oscillates with a period of 3.2\,mm ($2\pi/\Delta\kappa$) and peaks at 0.05\,\%. The near-axis electron density behind the drive stays within 0.87--1.10 of the background, and no electrons are trapped. With the density raised by 1\,\% at the same spot size, the wake is unchanged and the phase slope rises by 0.020\,rad/mm (predicted 0.022); with the spot size raised by 1\,\% at the same density, it rises by 0.026\,rad/mm (predicted 0.033--0.021 for the four pairs, mean 0.027). Both independent sensitivities are thus confirmed. Retaining the azimuthal mode $m = 2$ changes none of these numbers in the first three digits (Methods). Over 10\,mm the same drive sustains 16--20.5\,GV/m with a phase slope of 0.047\,rad/mm, and a test electron at fixed phase gains 181\,MeV, still growing; the same comb in LG$_0$ reaches 27\,MeV at 2.0\,mm, its wake falls to 2.5\,GV/m and its phase moves at 2.05\,rad/mm. Energy flows down the ladder faster than at $3\times10^{18}$\,cm$^{-3}$: the top line falls from 0.29 to 0.13 and the lowest rises from 0.04 to 0.19 of the drive energy, and the Stokes line peaks at 0.2\,\%. The drive loses 5.7\,\% of its field energy, of which 94\,\% appears as wake energy (99\,\% at $3\times10^{18}$\,cm$^{-3}$). The drive energy is counted in a window that follows the drive envelope, which slips back by $c\kappa/\omega \approx 3.5\,\mu$m/mm; a window fixed at the initial position loses the tail of the drive after 6--8\,mm and overestimates the loss.

\paragraph{9.8\,mJ over 30\,mm.} The same ladder with all amplitudes reduced by $\sqrt{3}$ (9.8\,mJ, linear-theory wake 7.5\,GV/m) holds 6.2\,GV/m (0.83--0.87 of linear theory), flat to 20\,mm and 10\,\% lower at 30\,mm, with a phase slope of 0.049\,rad/mm (1.5\,rad over 30\,mm); a test electron at the best fixed phase gains 169\,MeV (about 185\,MeV without the drift). The drive envelope slips back by 105\,$\mu$m over the 30\,mm; the wake window and the drive-energy cut follow it (Methods), and a fixed window gives amplitudes within 7\,\% of these, so neither the decline at 30\,mm nor the ratio to linear theory is a window artefact. A real bunch would have to start at least 105\,$\mu$m behind the drive tail, where the wake is 0.85--0.94 of its value at the tail (wake far behind the drive, Section~S8), and would gain about 7\,\% less (estimate). The accumulated drift, 1.5\,rad, is as large as the accelerating-and-focusing window; since it is linear in $z$ and a 1\,\% width change shifts the slope by 0.026\,rad/mm (Section~S7 above), a width offset of 1--2\,\% should null it (projected). The drive loses 6.6\,\% of its energy, 81\,\% of which is found as wake energy, and transfers 5.4\,\% of its energy to the wake over the 30\,mm, as the 29\,mJ drive over 10\,mm. Its line energies at 30\,mm (0.22, 0.12, 0.18, 0.19, 0.13 of the initial drive energy from the lowest to the highest line) are those of the 29\,mJ drive at 10\,mm (0.22, 0.14, 0.18, 0.20, 0.13): the energy flow down the ladder proceeds at a rate proportional to the intensity, so the gain per stage, $\approx E_zL$, is nearly independent of the drive energy. The ratio to linear theory (0.83--0.87) is below the value expected from the amplitude scan of Section~S6 for this amplitude (about 0.93); the origin of this additional deficit in long stages is open. The drive-loss values quoted here and in the Methods are field-energy losses; the sum of the projections on the five lines falls faster, because it also counts energy that leaves these modes. A two-stage run and a beam-loading scan with this drive are in Section~S8.

\section{Further checks}
Checks reported elsewhere: the amplitude scan of the deficit, the seven-line luminal ladder and the ladder at $10^{19}$\,cm$^{-3}$ (Section~S6), the azimuthal mode $m = 2$ (Methods) and the 1\,\% spot-size error (Section~S7).

\paragraph{Full-length convergence.} For the 10\,mm run at $6\times10^{18}$\,cm$^{-3}$ (Section~S7), with $\Delta r = 0.25\,\mu$m and $\Delta\zeta = \lambda/36$ the wake amplitude at every $z$, the test-electron gain (181\,MeV in both), the drive loss (5.65 against 5.67\,\%), the energies of the lowest and highest line and the Stokes energy agree to within 1\,\%; the phase slope falls from 0.047 to 0.032\,rad/mm, i.e.\ about a third of the residual drift is numerical dispersion.

\paragraph{A ladder seeded with two lines.} With only LG$_0$ and LG$_1$ seeded (23\,mJ) and the upper rungs at $a = 10^{-4}$, the ladder does not build itself over 10\,mm: energy flows down from LG$_1$ into LG$_0$ (0.54 to 0.03 and 0.46 to 0.68 of the drive energy), LG$_2$ grows only to 1.1\,\% and LG$_3$ to 0.05\,\%. The wake starts at 13.6\,GV/m (0.94 of the two-line linear estimate) and decays with the depleting LG$_1$ to about 6\,GV/m; a test electron gains 78\,MeV; the drive loses only 3.6\,\% of its energy. All rungs must therefore be seeded.

\paragraph{The wake far behind the drive.} The $k_p$ component of $E_z$ on the axis behind the 29\,mJ drive at $6\times10^{18}$\,cm$^{-3}$, evaluated over windows of two plasma wavelengths, is stationary between 0.5 and 3\,mm. Relative to its value at the drive tail (19.7\,GV/m) it is 0.94 at 57\,$\mu$m, 0.85 at 112\,$\mu$m, 0.76 at 139\,$\mu$m and 0.61 at 166\,$\mu$m behind the tail, and erratic beyond 190\,$\mu$m. The wave fronts curve backwards, because the plasma frequency rises off axis in the channel: at $r = 10\,\mu$m the phase lags the axis by 1.1\,rad at the tail and 1.8\,rad 140\,$\mu$m behind. The phase of the focusing field $W_r = E_r - cB_\theta$ relative to $E_z$ (from the Panofsky--Wenzel theorem) grows from 1.9\,rad at the tail to 2.3 and 2.5\,rad at 140 and 166\,$\mu$m (an ideal wake: $\pi/2$), so the accelerating-and-focusing window shrinks from 1.57\,rad to about 1.2, 0.9 and 0.6\,rad. The wake is therefore usable up to about 140\,$\mu$m behind the drive tail (at least 0.76 of the gradient, an accelerating-and-focusing window of at least 0.8\,rad).

\paragraph{Beam loading.} A Gaussian witness ($\sigma_z = \sigma_r = 1\,\mu$m, 100\,MeV) at the accelerating-and-focusing phase of the 29\,mJ ladder at $6\times10^{18}$\,cm$^{-3}$, over 1\,mm (Table~\ref{tab:loading}). The unloaded wake carries 0.158\,mJ per millimetre. The gain falls linearly with the charge, $\approx 14.2 - 1.53\,Q[\mathrm{pC}]$\,MeV/mm, so the extracted energy peaks near 5\,pC: 32\,$\mu$J/mm, 20\,\% of the wake energy and 0.11\,\% of the laser energy per millimetre. Loading also flattens the field slope over the bunch, so 5\,pC gives the smallest energy spread; 10\,pC overloads the wake and the bunch tail is decelerated. The drop of the wake energy behind the witness agrees with the extracted energy to about 20\,\%. The 2\,pC case reproduces the first millimetre of the two-stage run below to all printed digits. A 5\,pC witness with a current falling linearly from head to tail, designed for a flat loaded field from the measured loading kernel, gains 7.2\,MeV/mm instead of 6.5 and reaches an rms energy spread of 1.09\,\% instead of 1.73\,\% after 1\,mm (90\,\% energy width 3.8 against 5.2\,MeV) at the same emittance; it extracts 36\,$\mu$J/mm, 23\,\% of the wake energy (0.12\,\% of the laser energy per millimetre). The field is not yet flat (6.6\,MeV/mm in the tail quarter, 8.5 in the head quarter): the witness focuses from 1 to about 0.4\,$\mu$m within 0.3\,mm, which raises its loading above the kernel measured with the Gaussian bunch.

\begin{table}[H]
\centering
\begin{tabular}{cccccc}
\toprule
$Q$ [pC] & gain [MeV/mm] & $\Delta E/E$ & $\varepsilon_n$ [$\mu$m] & extracted [$\mu$J/mm] & of wake energy \\
\midrule
0.5 & 13.4 & 3.2\,\% & 0.53/0.49 & 6.7 & 4\,\% \\
1 & 12.7 & 2.9\,\% & 0.51/0.49 & 12.7 & 8\,\% \\
2 & 11.2 & 2.4\,\% & 0.48/0.46 & 22.4 & 14\,\% \\
5 & 6.5 & 1.7\,\% & 0.52/0.51 & 32.4 & 20\,\% \\
10 & $-1.3$ & 5.1\,\% & 0.63/0.63 & net loss & -- \\
\bottomrule
\end{tabular}
\caption{Beam loading of the 29\,mJ ladder wake at $6\times10^{18}$\,cm$^{-3}$ (Gaussian witness, initial $\varepsilon_n = 0.3\,\mu$m; energy spread and emittance after 1\,mm).}
\label{tab:loading}
\end{table}

\paragraph{Two stages with a re-used drive.} Stage 1 is the 10\,mm run of Section~S7 with a 2\,pC witness; the drive is not changed by the witness (its line energies equal those of the run without a witness to all printed digits). For stage 2 the five lines leaving stage 1 are re-launched in their modes with their measured energies (25.4\,mJ, 0.86 of the original drive; the energy outside the five lines and the envelope distortion are dropped) and the witness is read back at the same phase (Table~\ref{tab:stages}). Stage 2 gains 67\,MeV, two thirds of stage 1: the energy that flowed down the ladder in stage 1 is missing from the upper rungs, and it keeps flowing, so LG$_1$ falls from 0.157 to 0.037 of the stage-2 drive energy over 10\,mm and the wake decays from 12.4 to about 8\,GV/m. The witness gains about half of the wake amplitude because 2\,pC loads the wake by about 3.5\,GV/m and the residual phase drift moves the wake crest towards the bunch; the 16--19\,\% energy spread is set by the wake curvature over $k_p\sigma_z = 0.46$ and the loading slope, and the emittance grows because the 1\,$\mu$m beam is not matched to the linear focusing. Laser-to-witness efficiency: 0.66\,\% in stage 1, 0.53\,\% in stage 2 (of the re-launched 25.4\,mJ), 1.12\,\% of the 29.4\,mJ laser in total; wake plus witness account for 96\,\% of the drive loss in both stages. A second run of stage 2 with a drive whose pair sum $\sum_p a_pa_{p+1}$ was 6.2\,\% weaker (23.7\,mJ, mostly less energy in the lowest line) ends at 263.8 instead of 264.9\,MeV: the stronger drive gains 6\,\% more in the first millimetres, but its stronger wake also pumps LG$_1$ down into LG$_0$ faster (LG$_1$ at 10\,mm: 0.037 against 0.047), so the beat strength $\sum_q\sqrt{\mathcal{E}_q\mathcal{E}_{q+1}}$ is equal in both runs from about 5\,mm on, and the extra energy ends in the lowest line, where it drives nothing once LG$_1$ is empty. Re-using a depleted drive is thus limited by the cascade itself: a stronger second stage needs more energy in the upper lines, not in the lowest one.

\begin{table}[H]
\centering
\begin{tabular}{lccc}
\toprule
 & start & after stage 1 (10\,mm) & after stage 2 (20\,mm) \\
\midrule
energy & 100\,MeV & 198\,MeV & 265\,MeV \\
$\Delta E/E$ (rms) & 0.1\,\% & 16\,\% & 18.5\,\% \\
$\varepsilon_n$ ($x/y$) & 0.30/0.30\,$\mu$m & 0.53/0.53\,$\mu$m & 0.78/0.85\,$\mu$m \\
charge & 2.000\,pC & 1.998\,pC & 1.998\,pC \\
drive energy & 29.4\,mJ & 27.7\,mJ (field), 25.4\,mJ re-launched & 24.4\,mJ \\
\bottomrule
\end{tabular}
\caption{Two 10\,mm stages at $6\times10^{18}$\,cm$^{-3}$ with a 2\,pC witness and the drive of stage 1 re-launched in stage 2.}
\label{tab:stages}
\end{table}

% ------------------------------------------------------------------------------------------------
\section{Metal capillaries}
A linearly polarized $J_0$ profile is not a mode of a conducting guide: it splits into TE$_{1n}$ and TM$_{1n}$ modes that dephase, and a two-mode wake collapses after 0.4\,mm in our simulations; an analytic TE/TM decomposition reproduces the purity decay to 1--2\,\%. $J_0$ is the EH$_{1n}$ mode of a dielectric or plasma capillary. The ladder therefore needs plasma channels or dielectric capillaries, not metal guides. (E. A. J. Marcatili and R. A. Schmeltzer, Bell Syst. Tech. J. \textbf{43}, 1783 (1964).)

% ------------------------------------------------------------------------------------------------
\section{Projection onto Gaussian channels}
We compute the guided $m = 0$ modes of $n = n_\infty[1 - D\exp(-r^2/s^2)]$ (on-axis density $n_0 = (1-D)n_\infty$) by solving $[-r^{-1}\partial_r(r\partial_r) + k_{p0}^2(n - n_0)/n_0]u = k_\perp^2u$; a mode is guided if $k_\perp^2 < k_{p0}^2D/(1-D)$. For each $D$ and number of lines $N$ the width $s$ is chosen such that the ladder ($\kappa$ from the mean eigenvalue spacing of the first $N$ modes, lines on their modes) has its central line at 1.03\,$\mu$m. Table~\ref{tab:gauss} gives the largest deviation of the eigenvalue spacings from their mean. The beats remain resonant if this deviation is below about $1/(2N_\tau)$ (4.5\,\% for 0.5\,ps at $6\times10^{18}$\,cm$^{-3}$, twice that for 0.25\,ps). These are projections from the mode spectrum; particle-in-cell simulations in Gaussian channels are left for future work.

\begin{table}[H]
\centering
\small\setlength{\tabcolsep}{4pt}
\begin{tabular}{lcccc}
\toprule
depletion $D$ & 2 lines & 3 lines & 4 lines & 5 lines \\
\midrule
20\,\% & uniform ($s = 42\,\mu$m) & 22\,\%, top mode at the guiding edge & not guided & not guided \\
30\,\% & uniform & 5.5\,\% ($s = 57\,\mu$m) & 18\,\% & not guided \\
40\,\% & uniform & 3.0\,\% ($s = 76\,\mu$m) & 7.6\,\% ($s = 66\,\mu$m) & 13\,\% \\
\bottomrule
\end{tabular}
\caption{Spacing non-uniformity of the guided modes of Gaussian channels at $n_0 = 6\times10^{18}$\,cm$^{-3}$, with the channel width $s$ that places the central line at 1.03\,$\mu$m. At $10^{19}$\,cm$^{-3}$ the numbers are similar (three lines: 9\,\% at $D = 30\,\%$, 4.5\,\% at 40\,\%).}
\label{tab:gauss}
\end{table}

The Manley--Rowe limit of the laser-to-wake efficiency falls with the number of lines: about 6--8\,\% for two, 10--13\,\% for three and 14--18\,\% for four lines at $6\times10^{18}$--$10^{19}$\,cm$^{-3}$. Non-consecutive modes with equal gaps (e.g.\ LG$_0$, LG$_1$, LG$_3$ in a saturating channel) offer a further option that we have not explored.


\begin{thebibliography}{99}
\small
\bibitem{Tajima1979} T. Tajima and J. M. Dawson, Laser electron accelerator, Phys. Rev. Lett. \textbf{43}, 267 (1979).
\bibitem{Esarey2009} E. Esarey, C. B. Schroeder and W. P. Leemans, Physics of laser-driven plasma-based electron accelerators, Rev. Mod. Phys. \textbf{81}, 1229 (2009).
\bibitem{Gonsalves2019} A. J. Gonsalves \textit{et al.}, Petawatt laser guiding and electron beam acceleration to 8\,GeV in a laser-heated capillary discharge waveguide, Phys. Rev. Lett. \textbf{122}, 084801 (2019).
\bibitem{Picksley2024} A. Picksley \textit{et al.}, Matched guiding and controlled injection in dark-current-free, 10-GeV-class, channel-guided laser plasma accelerators, Phys. Rev. Lett. \textbf{133}, 255001 (2024).
\bibitem{AWAKE2018} AWAKE Collaboration, Acceleration of electrons in the plasma wakefield of a proton bunch, Nature \textbf{561}, 363 (2018).
\bibitem{Nubbemeyer2017} T. Nubbemeyer \textit{et al.}, 1\,kW, 200\,mJ picosecond thin-disk laser system, Opt. Lett. \textbf{42}, 1381 (2017).
\bibitem{Dira} TRUMPF Scientific Lasers, Dira 200-1 thin-disk amplifier (200\,mJ, 1\,kHz, 1030\,nm), product information (2026).
\bibitem{Faure2019} J. Faure \textit{et al.}, A review of recent progress on laser-plasma acceleration at kHz repetition rate, Plasma Phys. Control. Fusion \textbf{61}, 014012 (2019).
\bibitem{Salehi2021} F. Salehi \textit{et al.}, Laser-accelerated, low-divergence 15-MeV quasimonoenergetic electron bunches at 1\,kHz, Phys. Rev. X \textbf{11}, 021055 (2021).
\bibitem{Lazzarini2024} C. M. Lazzarini \textit{et al.}, Ultrarelativistic electron beams accelerated by terawatt scalable kHz laser, Phys. Plasmas \textbf{31}, 030703 (2024).
\bibitem{Rosenbluth1972} M. N. Rosenbluth and C. S. Liu, Excitation of plasma waves by two laser beams, Phys. Rev. Lett. \textbf{29}, 701 (1972).
\bibitem{Clayton1993} C. E. Clayton \textit{et al.}, Ultrahigh-gradient acceleration of injected electrons by laser-excited relativistic electron plasma waves, Phys. Rev. Lett. \textbf{70}, 37 (1993).
\bibitem{Hooker2014} S. M. Hooker \textit{et al.}, Multi-pulse laser wakefield acceleration: a new route to efficient, high-repetition-rate plasma accelerators and high flux radiation sources, J. Phys. B \textbf{47}, 234003 (2014).
\bibitem{Cowley2017} J. Cowley \textit{et al.}, Excitation and control of plasma wakefields by multiple laser pulses, Phys. Rev. Lett. \textbf{119}, 044802 (2017).
\bibitem{Jakobsson2021} O. Jakobsson, S. M. Hooker and R. Walczak, GeV-scale accelerators driven by plasma-modulated pulses from kilohertz lasers, Phys. Rev. Lett. \textbf{127}, 184801 (2021).
\bibitem{Ross2023} A. J. Ross \textit{et al.}, Resonant excitation of plasma waves in a plasma channel, Phys. Rev. Research \textbf{6}, L022001 (2024).
\bibitem{vandeWetering2023} J. J. van de Wetering, S. M. Hooker and R. Walczak, Stability of the modulator in a plasma-modulated plasma accelerator, Phys. Rev. E \textbf{108}, 015204 (2023).
\bibitem{Froula2018} D. H. Froula \textit{et al.}, Spatiotemporal control of laser intensity, Nat. Photon. \textbf{12}, 262 (2018).
\bibitem{Palastro2020} J. P. Palastro \textit{et al.}, Dephasingless laser wakefield acceleration, Phys. Rev. Lett. \textbf{124}, 134802 (2020).
\bibitem{Caizergues2020} C. Caizergues \textit{et al.}, Phase-locked laser-wakefield electron acceleration, Nat. Photon. \textbf{14}, 475 (2020).
\bibitem{Smartsev2019} S. Smartsev \textit{et al.}, Axiparabola: a long-focal-depth, high-resolution mirror for broadband high-intensity lasers, Opt. Lett. \textbf{44}, 3414 (2019).
\bibitem{Debus2019} A. Debus \textit{et al.}, Circumventing the dephasing and depletion limits of laser-wakefield acceleration, Phys. Rev. X \textbf{9}, 031044 (2019).
\bibitem{Palastro2025} J. P. Palastro, K. G. Miller, M. R. Edwards, A. L. Elliott, L. S. Mack, D. Singh and A. G. R. Thomas, Arbitrary-velocity laser pulses in plasma waveguides, Phys. Rev. Research \textbf{7}, 023249 (2025).
\bibitem{Palastro2026} J. P. Palastro \textit{et al.}, Dephasingless laser wakefield acceleration in a plasma waveguide, Phys. Rev. Research \textbf{8}, 033364 (2026).
\bibitem{Pukhov2023} A. Pukhov, N. E. Andreev, A. A. Golovanov, I. I. Artemenko and I. Yu. Kostyukov, Laser--plasma wake velocity control by multi-mode beatwave excitation in a channel, Plasma \textbf{6}, 29 (2023).
\bibitem{GEMPICRZ_paper} A. Pukhov, GEM-PIC-RZ: a Galilean electromagnetic particle-in-cell code in quasi-cylindrical geometry with nonuniform grids, arXiv:2610.31622 (2026).
\bibitem{GEMPICRZ} A. Pukhov, GEM-PIC-RZ, \url{https://github.com/AlexMPukhov/GEM-PIC-RZ} (2026), doi:10.5281/zenodo.23600179.
\bibitem{GEMPIC} A. Pukhov, N. Elkina and T. Wilson, Galilean electromagnetic particle-in-cell code, J. Comput. Phys. \textbf{567}, 115331 (2026).
\bibitem{Sokolov2000} A. V. Sokolov, D. R. Walker, D. D. Yavuz, G. Y. Yin and S. E. Harris, Raman generation by phased and antiphased molecular states, Phys. Rev. Lett. \textbf{85}, 562 (2000).
\bibitem{Alejo2022} A. Alejo, J. Cowley, A. Picksley, R. Walczak and S. M. Hooker, Demonstration of kilohertz operation of hydrodynamic optical-field-ionized plasma channels, Phys. Rev. Accel. Beams \textbf{25}, 011301 (2022).
\bibitem{Shalloo2018} R. J. Shalloo \textit{et al.}, Hydrodynamic optical-field-ionized plasma channels, Phys. Rev. E \textbf{97}, 053203 (2018).
\bibitem{Picksley2020} A. Picksley \textit{et al.}, Meter-scale conditioned hydrodynamic optical-field-ionized plasma channels, Phys. Rev. E \textbf{102}, 053201 (2020).
\bibitem{Oubrerie2022} K. Oubrerie \textit{et al.}, Controlled acceleration of GeV electron beams in an all-optical plasma waveguide, Light Sci. Appl. \textbf{11}, 180 (2022).
\bibitem{Claude} Anthropic, Claude, large language model, \url{https://www.anthropic.com/claude} (2026).
\end{thebibliography}
\end{document}